\documentclass[]{article}

\usepackage{graphicx}
\usepackage{proof}
\usepackage{xcolor}
\usepackage{amssymb}
\usepackage{amsmath}
 \usepackage{amsthm}
 \newtheorem{proposition}{Proposition}
 \newtheorem{theorem}{Theorem}
 \newtheorem{lemma}{Lemma}
 \theoremstyle{definition}
  \newtheorem{example}{Example}
 \newtheorem{definition}{Definition}
\usepackage{stmaryrd} 
\usepackage{colonequals}
\usepackage{xspace}
\usepackage{bm}
\usepackage{hyperref}

\newcommand{\mypar}[1]{\smallskip\noindent{\bf #1}}

\definecolor{brickred}{rgb}{0.8, 0.25, 0.33}
\newcommand{\bhl}[1]{{\color{blue}{#1}}} 
\newcommand{\phl}[1]{{\color{brickred}{#1}}} 
\newcommand{\thl}[1]{#1} 

\newcommand{\tcost}{\mathbf{cost}}
\newcommand{\tname}{\mathbf{name}}

\newcommand{\taddr}{\mathbf{addr}}

\newcommand{\msf}[1]{\mathsf{#1}}
\newcommand{\tsc}[1]{\textsc{#1}}

\newcommand*\mdelim[3]{%
	\mathopen{}\left#1%
	#3%
	\right#2\mathclose{}%
}
\newcommand*{\set}[2][]{\mdelim{\lbrace}{\rbrace_{#1}}{#2}}
\newcommand*{\ctx}[1]{\mdelim{\lbrace}{\rbrace}{#1}}
\renewcommand{\emptyset}{\varnothing}
\newcommand{\reducesto}{\longrightarrow}
\newcommand{\lts}[1]{\xrightarrow{{#1}}}
\newcommand{\enc}[1]{\llcorner {#1}\lrcorner}

\newcommand{\pp}{{\ \;\boldsymbol{|}\ \;}}
\newcommand{\res}[1]{(\boldsymbol\nu #1)\,}
\newcommand{\fwd}[2]{\phl{#1\leftrightarrow #2}}
\newcommand{\wait}[2]{\phl{#1().#2}}
\newcommand{\close}[1]{\phl{{#1}[]}}
\newcommand{\send}[4]{\phl{{#1}[#2\triangleright #3].#4}}
\newcommand{\recv}[3]{\phl{{#1}(#2).#3}}
\newcommand{\Case}[3]{\phl{#1.\msf{case}(#2,#3)}}
\newcommand{\inl}[2]{\phl{{#1}[\msf{inl}].#2}}
\newcommand{\inr}[2]{\phl{{#1}[\msf{inr}].#2}}
\newcommand{\srv}[3]{\phl{\bang #1(#2).#3}}
\newcommand{\client}[3]{\phl{\query #1[#2].#3}}

\newcommand{\dual}[1]{#1^{\bot}} 
\newcommand{\one}{\mathbf{1}}
\newcommand{\tensor}{\otimes}
\newcommand{\parr}{\mathbin{\bindnasrepma}}
\newcommand{\with}{\mathbin{\binampersand}}
\newcommand{\bang}{\mathord{!}}
\newcommand{\query}{\mathord{?}}

\newcommand{\seq}{\Vdash}  
\newcommand{\cll}{\vdash}  
\newcommand{\coherence}{\vDash}  
\newcommand{\br}[2][]{[^{\phl{#1}} #2]}
\newcommand{\dbr}[1]{[\![#1]\!]}
\newcommand{\pairQ}[2] { \phl{#1}:\thl{#2}}
\newcommand{\Star}{\textcolor{blue}{\ast}}
\newcommand{\Query}{\textcolor{blue}{\mathcal Q}}
\newcommand{\Left}{\textcolor{blue}{\mathcal L}} 
\newcommand{\Right}{\textcolor{blue}{\mathcal R}} 
\newcommand{\Symbol}{\textcolor{blue}{\mathcal{X}}}

\newcommand{\cut}{\tsc{Cut}\xspace}

\newdimen\mydisplayskip
\newenvironment{smallequation}
{\par\nobreak\vskip\mydisplayskip\noindent\bgroup\small\csname equation\endcsname}{\csname endequation\endcsname\egroup}
\newenvironment{smallequation*}
{\par\nobreak\vskip\mydisplayskip\noindent\bgroup\small\csname equation*\endcsname}{\csname endequation*\endcsname\egroup}

\newenvironment{smallalign*}
{\par\nobreak\noindent\bgroup\small\csname align*\endcsname}{\csname endalign*\endcsname\egroup}

\newcommand\mydots{\hbox to 1em{.\hss.\hss.}}

\begin{document}
\title{Forwarders as Process Compatibility, Logically}
%
%
\author{Marco Carbone
	\and Sonia Marin
	\and Carsten Sch\"urmann
}
%
%
%
\maketitle              
\begin{abstract}
  Session types define protocols that processes must follow when
  communicating.  The special case of binary session types, 
  i.e.\ type annotations of protocols between two parties, is known to be in
  a pro\-po\-si\-tions-as-types correspondence with linear logic. In
  previous work, we have shown that the generalization to multiparty
  session types can be expressed either by coherence proofs or by
  arbiters, processes that act as middleware by forwarding messages
  according to the given protocol. In this paper, following the
  propositions-as-types fashion, we generalize arbiters to a logic,
  which we call forwarder logic, a fragment of classical linear logic
  still satisfying cut-elimination. Our main result is summarized as
  follows: forwarders generalize coherence and give an elegant
  proof-theoretic characterization of multiparty compatibility, a
  property of concurrent systems guaranteeing that all sent messages
  are eventually received and no deadlock ever occurs.

\end{abstract}

\section{Introduction}\label{sec:introduction}
A concurrent system is more than a sum of processes. It also includes
the fabric that determines how processes are tied together.
%
Session types, originally proposed by Honda et al.~\cite{HVK98}, are
type annotations that ascribe protocols to processes in a concurrent
system and 
determine how they behave when communicating
with each other. Such type annotations are useful for various
reasons. First, they serve as communication blueprints for the entire
system and give programmers clear guidance on how to implement
communication patterns at each endpoint (process or service).  Second,
they make implementations of concurrent systems safer, since
well-typedness entails basic safety properties of programs such as
{\em lack of communication errors} (``if the protocol says I should
receive, e.g., an integer, I will never receive, e.g., a boolean''),
{\em session fidelity} (``my programs follow the protocol
specification patterns''), and {\em in-session deadlock freedom}
(``the system never gets stuck by running a protocol'').
%
Intuitively, session types make sure that the processes are {\em compatible}
and that they exchange messages in the prescribed way for the
concurrent system to work correctly.
They do that by preventing messages from being duplicated, as
superfluous messages would not be accounted for, and by preventing
messages from getting lost, otherwise a process might get stuck,
awaiting a message.

In the case of \emph{binary sessions types}, type annotations of
protocols between two parties, 
compatibility means for 
type annotations to be dual to one another: the send action of one
party must be matched by a corresponding receive action of the other
party, 
and vice versa.  Curiously, binary session types find their logical
foundations in linear logic, as identified by Caires and
Pfenning~\cite{CP10,CP16} and later by Wadler~\cite{W12,W14}.  They
have shown that session types correspond to linear logic propositions,
processes to proofs, reductions in the operational semantics to cut
reductions in linear logic proofs, and compatibility to the logical
notion of duality for linear formulas.  Duality, thus, defines, for
the lack of a better word, the ``fabric'' through which two processes
communicate 
while abstracting away from practical details, e.g., message delay,
message order, or message buffering.

The situation is not as direct for \emph{multiparty session
  types}~\cite{HYC08,HYC16}, 
type annotations for protocols with more than two participants.
Carbone et al.~\cite{CMSY15,CLMSW16} extended Wadler's embedding of
binary session types into classical linear logic (CLL) to the
multiparty setting, by generalising duality to the notion of {\em
  coherence}.
They observed that the in-between fabric, through which multiple processes communicate, holds the very key to
understanding multiparty session types:
when forcing the type annotations to be {\em coherent}, one ensures
that sent messages will eventually be collected.
%
Coherence as a deductive system allows one to derive  
compatible judgements, while proofs correspond precisely to multiparty
protocol specifications. 
A key result is that coherence proofs can be encoded as well-typed (as
proofs in CLL) processes, called {\em arbiters}, which means that the
fabric can be modelled formally as a process-in-the-middle.
However, no precise logical characterisation of what constitutes
arbiters was given.
In this paper, we continue this line of research and define a
subsystem of processes, called \emph{asynchronous forwarders} or \emph{forwarders} in short, that
provides one possible such characterisation and also  guarantees multiparty session
compatibility.
%

As the name already suggests, a forwarder is a process that forwards
messages, choices, and services from one endpoint to another according
to the protocol specification.
Intuitively, similarly to an arbiter, a forwarder process mimics the
fabric by capturing the message flow.  However, when data-dependencies
allow, forwarders could, in theory, non-deterministically choose to
receive messages from different endpoints, and then forward such
messages at a later point. Or, they can also decide to buffer a
certain number of messages from a given receiver.  Eventually,
they
re-transmit messages only after receiving them,
without interpreting, modifying, or computing with them.

In this work, asynchronous forwarders support buffers of unlimited
size.
This  preserves the order of messages from the same sender, i.e., after
receiving a message from one party, the forwarder enqueues it until the message is delivered to its
destination.
%
Forwarders can be used to explain communication patterns as they
occur in practice, such as message routing, proxy services, and
runtime monitors for message flows~\cite{JGP16}.
%

The meta-theoretic study of forwarders allows us to
conclude that there is a correspondence between forwarders and multiparty compatibility.
Forwarders are stronger than coherence and can be used to guide the communication of multiple processes similarly to the multi-cut elimination in~\cite{CLMSW16} and so they are a way to justify multiparty compatibility.
The reverse direction also holds, i.e., if a multiparty session of processes is compatible, they can be linked by a forwarder; in such a way, forwarders as a logic provide a proof theory for multiparty compatibility.
In this paper, we also show that 
forwarders can safely be composed through cut elimination, which
allows us to combine the fabric between two concurrent systems
(figuring arbitrary many processes). 
%

\smallskip

\mypar{Outline and key contributions.}  The key contributions of this
paper include
\begin{itemize}
\item a logical characterisation of \emph{forwarders}
  (\S~\ref{sec:forwarders});
\item a reductive operational semantics based on cut-elimination
  (\S~\ref{sec:semantics});
\item a correspondence between multiparty compatibility and 
forwarders, generalising coherence: every 
forwarder guarantees correct multiparty communication (\S~\ref{sec:mcut}) 
and any compatible multiparty session can be
emulated by a forwarder (\S~\ref{sec:mcompatibility}).
\end{itemize}
Additionally, \S~\ref{sec:preview} gently introduces the main concepts
on an example and \S~\ref{sec:prelim} recaps the definitions of types,
processes, and CP-typing, while \S~\ref{sec:related} discusses related
and future work and concluding remarks are in
\S~\ref{sec:conclusions}.

\section{Preview}\label{sec:preview}
We now proceed with a gentle introduction to asynchronous forwarders
by informally describing the
classic \emph{2-buyer protocol}~\cite{HYC08,HYC16}, where two buyers
intend to buy a book jointly from a seller.
The first buyer
sends the title of the book 
to the seller,
who, in turn, sends a quote to both buyers.
Then, 
the first buyer decides how much she wishes to contribute and informs
the second buyer,
who either pays the rest or cancels the transaction by informing the
seller.

The three participants 
are connected through endpoints $\phl{b_1}$, $\phl{b_2}$, and $\phl s$
respectively. Each endpoint must be used according to its respective
session type annotation 
which gives a precise description of how each endpoint has
to act. 
\begin{smallequation}\label{eq:cotypes}
  \begin{array}{c}
    {\phl{b_1}} : \tname \tensor \tcost^\perp \parr \tcost \tensor
    \one
    \quad
    \phl{b_2} : \tcost^\perp \parr \tcost^\perp \parr ((\taddr \tensor \one) \oplus \one)
    \\[1mm]
    \phl{s} : \tname^\perp \parr \tcost \tensor \tcost
    \tensor ((\taddr^\perp \parr \bot) \with \bot)
  \end{array}
\end{smallequation}
For example,
$\phl{b_1}: \tname \tensor \tcost^\perp \parr \tcost \tensor \one$ says that
buyer $\phl{b_1}$ must first send a value of type $\tname$ (the title
of the book), then receive a value of type $\tcost$ (the price of the
book), then send a value of type $\tcost$ (the amount of money she
wishes to contribute), and finally terminate.

Any three processes 
who respectively
use endpoints $\phl{b_1}$, $\phl{b_2}$, and~$\phl s$ according to the
type specifications above are going to execute this protocol
correctly because these type specifications are \emph{compatible}. 
In a binary setting (only two endpoints communicating to
each other) compatibility is usually expressed by type duality: the
dual of a type is the type obtained by inverting every output with an
input, and vice-versa. 
In a multiparty setting, such compatibility can
be expressed as \emph{coherence}~\cite{CLMSW16} but, we will argue, also with asynchronous
forwarders.
The core idea of multiparty compatibility is to establish for each output which endpoint should
receive it.
We propose to base compatibility on whether there exists a process (\emph{forwarder}) able to glue the (duals of the) types of the different endpoints,
e.g., in the 2-buyers case, we can establish compatibility if
there is a process typable in the context formed by the duals of the types in~\eqref{eq:cotypes}. 
Indeed, such a forwarder process  exists and could have the following simple behaviour:
the request containing the book name is received over
some endpoint connected to $\phl{b_1}$ and then forwarded over an
endpoint connected to $\phl s$, then the same is done over
$\phl{b_1}$ and $\phl{b_2}$ for the amount, and so on. 
We observe that such a
process, provided that it indeed respects the dual of the types in~\eqref{eq:cotypes},
could still have many different variations. For example, the first {send} 
can happen at a later point rather than immediately after the request 
has been {received}. 
Yet, a forwarder cannot be any 
process: it must be such that i) anything that has been received is
eventually sent, ii) anything that is sent must have been previously received, 
and iii) the order of messages between any two points must be
preserved. 
Our theory of forwarders captures precisely such requirements.

Although the notion of coherence also satisfies these properties, 
by capturing all and only requirements i), ii), and iii), 
we can model the composition of processes that
cannot be captured by coherence. Consider for example 
two endpoints $\phl x$ and $\phl y$ willing to communicate with the
following protocol -- called a \emph{criss-cross}: they both send a message to each other, and then
the messages are received, according to the following types
\begin{smallequation*}
\pairQ x{\tname\tensor \tcost \parr \one}
\qquad\qquad 
\pairQ y{\tcost^\perp\tensor  \tname^\perp\parr \bot}
\end{smallequation*}
Such protocol leads to no error (assuming processes implement an
asynchronous semantics), still the two types above are not
coherent~\cite{CLMSW16}. On the other hand, we can easily write a
forwarder 
typable in the context
$\pairQ x{\tname^\perp\parr \tcost ^\perp\tensor \bot}, \pairQ
y{\tcost\parr \tname\tensor \one}$ formed by their duals, i.e., a
process that first receives on both $\phl x$ and $\phl y$ and then
forwards the received messages over to $\phl y$ and $\phl x$,
respectively.



%
%
%

%
%

\section{Preliminaries: CP and Classical Linear Logic}\label{sec:prelim}
In order to make the presentation of asynchronous forwarders easier to
comprehend, we give an introduction to the proposition-as-sessions
approach~\cite{W14}. This comprises
the syntax of types and processes and the interpretation of processes
as sequent proofs in classical linear logic (CLL). In the interest of space,
we restrict this presentation to the multiplicative fragment. The treatment
of the additive and exponential fragments can be found in the
appendix. 

\mypar{Types.} Following the propositions-as-types approach, types,
taken to be propositions (formulas) of CLL, are associated to names,
denoting the way an endpoint must be used at runtime. Their formal
syntax is given as:
\begin{smallequation}\label{eq:types}
	\thl{A} ::=  \quad
	\thl{a}
	\: \mid\: \thl{a^\perp}
	\: \mid\: \thl{1}
	\: \mid\: \thl{\bot}
	\: \mid\: \thl{(A\tensor A)}
	\: \mid\: \thl{(A\parr A)}
\end{smallequation}
Atoms $a$ and negated atoms $\dual a$ are basic dual types. 
Types $\one$ and
$\perp$ denote an endpoint that must close with a last
synchronisation. A type $A \tensor
B$ is assigned to an endpoint that outputs a message of type
$A$ and then is used as $B$. Similarly, an endpoint of type $A\parr
B$, receives a message of type $A$ and continues as
$B$.

\mypar{Duality.} Operators can be grouped in pairs of duals that reflect the
input-output duality. Consequently, standard duality
$(\cdot)^\perp$ on types is inductively defined as:
\begin{smallequation*}
	\dual{(\dual{a})} = a
	\qquad \dual\one=\perp
	\qquad \dual{(A\tensor B)} = \dual A\parr\dual B
\end{smallequation*}

\mypar{Processes.}  We use a standard language of {\em processes} to represent
communicating entities  (including forwarders) which is a variant of
the $\pi$-calculus~\cite{MPW92} with specific communication primitives
as usually done for session calculi. Moreover, given that the theory
of this paper is based on the proposition-as-sessions correspondence
with CLL, we adopt a syntax akin to that of
Wadler~\cite{W14}: 
%
\begin{smallequation*}
  \begin{array}{l@{\quad}l@{\quad}l@{\qquad}l@{\quad}l@{\quad}l@{\qquad}l}
    \phl P,\phl Q ::= & \fwd xy          & \text{(link)}
    &
      \phl{\res {xy} (P\!\!\pp\!\! Q)} & \text{(parallel)}
    \\
                      & \wait x P & \text{(wait)}
    &
      \close x & \text{(close)}
    \\
                      & \recv xyP        & \text{(input)}
                
    &
      \send xyPQ       & \text{(output)}
%
  \end{array}
\end{smallequation*}
%
A link $\fwd xy$ is a binary forwarder, i.e., a process that forwards
any communication between endpoints $\phl x$ and $\phl y$. This yields
a sort of equality relation on names: it says that endpoints $\phl x$
and $\phl y$ are equivalent, and communicating something over $\phl x$
is like communicating it over $\phl y$.
%
%
%
Note that we use endpoints instead of channels~\cite{V12}. The
difference is subtle: the restriction $\phl{\res {xy}}$ connects the
two endpoints $\phl x$ and $\phl y$, instead of referring to the
channel between them.
The terms $\wait xP$ and $\close x$ handle synchronisation (no message
passing); $\wait xP$ can be seen as an empty input on $\phl x$, while
$\close x$ terminates the execution of the process.
The term $\send xyPQ$ denotes a process that creates a fresh name
$\phl y$, spawns a new process $\phl P$, and then continues as
$\phl Q$. The intuition behind this communication operation is that
$\phl P$ uses $\phl y$ as an interface for dealing with the
continuation of the dual primitive (denoted by term $\recv xyR$, for
some $\phl R$).
We observe that Wadler~\cite{W14} uses the syntax $x[y].(P\,|\, Q)$,
but we believe that our version is more intuitive and gives a better
explanation of why we require two different processes to follow after
an output. However, our format is partially more restrictive, since
$\phl y$ is forced to be bound in $\phl P$ (which Wadler enforces with
typing). Also, note that output messages are always fresh, as for the
internal $\pi$-calculus~\cite{S96}, hence the output term $\send xyPQ$
is a compact version of the $\pi$-calculus term
$(\nu y)\, \overline x y.(P \,|\, Q)$.
%
%

\begin{figure}[t]
	\begin{smallequation*}
		\begin{array}{c}
			\infer[\textsc{Ax}]
			{\phl{\fwd{x}{y}} \cll \phl x:\thl {a^\bot}, \phl y:\thl{a}}
			{ }
			\qquad
			\infer[\bot]
			{\phl {\wait xP} \cll \thl \Delta, \phl x:\thl\bot}
			{\phl P \cll \thl\Delta}
			\qquad
			\infer[\one]
			{\phl{\close x} \cll \phl x: \thl\one}
			{ }
			\\[1ex]
			\infer[\tensor]
			{
				\phl{\send xyPQ} 
				\cll 
				\thl{\Delta_1}, \thl{\Delta_2}, \phl x:\thl{A_1 \tensor A_2}
			}
			{
				\phl P 
				\cll 
				\thl{\Delta_1}, \phl y:\thl A_1
				& \phl Q 
				\cll 
				\thl \Delta_2, \phl x: \thl A_2
			}
			\qquad\qquad\qquad
			\infer[\parr]
			{\phl{\recv xyP} 
				\cll \thl{\Delta}, \phl x: \thl {A_1 \parr A_2}
			}       
			{\phl P \cll \thl{\Delta}, \phl y:\thl A_1, \phl x:\thl A_2}
			\\[1ex]
		\end{array}
	\end{smallequation*}
	\caption{Sequent Calculus (Multiplicative Fragment) for CP and Classical Linear Logic} 
	\label{fig:cp}
\end{figure}

\mypar{CP-typing.}
As shown by Wadler~\cite{W14}, among all of the many process
expressions one can write, classical linear logic (CLL) characterises a subset that is
well-behaved, i.e.\ they satisfy deadlock freedom and session fidelity. 

Judgements are defined as $\phl{P} \cll \Delta$ with $\Delta$ a set of
named types, i.e.,
$\Delta \coloncolonequals \varnothing \mid \pairQ{x}{A}, \Delta$.
%
The system called CP, in Figure~\ref{fig:cp}, uses CLL to type processes.
%

CP can be extended with  a structural rule for defining composition of processes
which corresponds to the \textsc{Cut} rule from classical linear
logic:
\begin{smallequation*}
  \infer[\textsc{Cut}] {\phl{\res{xy}(P \pp Q)}
    \cll\thl\Sigma,\thl\Delta} {\phl P \cll\thl\Sigma, \phl x:\thl A &
    \phl Q \cll \thl\Delta, \phl y:\thl{\dual{A}}}
\end{smallequation*}
In linear logic this rule is admissible, i.e., 
the CLL derivations of the two premises can be combined into a derivation of the conclusion with no occurrence of the $\cut$ rule.
Moreover, this is a constructive procedure, called \emph{cut-elimination}, meaning that the proof with cut is inductively transformed into a proof without cut.
The strength of the proposition-as-type correspondence stems from the fact that it carries on to the proof level,
as it was shown that the cut-elimination steps correspond to reductions in the $\pi$-calculus~\cite{CP10,W12}.

%
%

\section{Asynchronous Forwarders}\label{sec:forwarders}
Following a proposition-as-types approach, we aim at a restriction of
CP such that derivable judgements are inhabited by forwarder processes
only. For the sake of clarity, we start our development focusing on
the multiplicative fragment of linear logic (rules $\textsc{Ax}$,
$\one$, $\perp$, $\tensor$, and $\parr$). We give a full extension to
additives and exponentials in the appendix.

Forwarders form a subclass of processes. Our development focuses
exclusively on those processes that are also typable in classical
linear logic. I.e., our goal is to identify all those CP processes
that are also forwarders. In order to do so, we must add further
information in the standard CP contexts.

\paragraph{Extended types.}
 To type forwarders, we extend the syntax of types (formulas) with
annotations that make explicit where messages should be forwarded from
and to. This is similar to local types $!{\mathsf p}.T$ and
$?{\mathsf p}.T$~\cite{CDYP15} expressing an output and an input to
and from role $\mathsf p$ respectively.
%
%
Intuitively, the reason for that is that in order to achieve cut
elimination for forwarders, we need to store more information on the
typing contexts (and endpoint types). We will see this in the section.
%
%
The meaning of each operator remains the same as in CP as shown in the
previous section.
The definition of duality is also the same.
%
\begin{smallequation*}
  \begin{array}{rclllll}
    \thl{B}  & \coloncolonequals &\quad
                                   \thl{a}
                                   \quad\mid\quad \thl{a^\perp} 
    &
      \mid\quad  \thl{\one^{\phl{u_1, \ldots, u_n}}}
    &
      \mid\quad \thl{\bot^{\phl u}} 
    &
      \mid\quad \thl{(A\tensor^{\phl u} B)} 
    &
      \mid\quad  \thl{(A\parr^{\phl u} B)} 
  \end{array}
\end{smallequation*}

The left hand side $A$ of $\tensor$ and $\parr$ are not annotated, as
in~\eqref{eq:types}, and become dynamically labelled when needed by
the typing routine. Note that in this section we only consider the
subset of $A$ without $\&$, $\oplus$ , $!$, and $?$.
We will see that units 
demonstrate some \emph{gathering} 
behaviour
%
which explains the need to annotate $\one$ 
with a non-empty list of an arbitrary number of distinct names.
We may write $\phl{\tilde u}$ for $\phl{u_1, \ldots, u_n}$ when the
size of the list is irrelevant.

We define a bidirectional map between annotated and non-annotated
types.
For a proposition $A$ defined as in~\eqref{eq:types}, we write $A(x)$
for the formula obtained by annotating every operator in $A$ with the
name $x$.
Conversely, we write $\enc{B}$ for the formula obtained from $B$ by
removing all the annotations.

\paragraph{Contexts.}
What we need is to be able to enforce the main features that
characterise a forwarder, namely i) any received message must be
forwarded, ii) any message that is going to be sent must be something
that has been previously received, and iii) the order of messages
between any two points must be preserved. In order to enforce these
requirements, we add more information to the standard CP
judgement. For example, let us consider the input process $\recv
xyP$. In CP, the typing environment for such process must be such that
endpoint $\phl x$ has type $A\parr B$ such that $\phl P$ has type
$\pairQ yA$ and $\pairQ xB$. However, the context is not telling us at
all that $\phl y$ is actually a message that has been received and, as
such, it should not be used by $\phl P$ for further communications but
just forwarded over some other channel. In order to remember this fact
when we type the subprocess $\phl P$, we actually insert $\pairQ yA$
into a queue that belongs to endpoint $\phl x$ where we put all the
types of messages received over it. I.e., when typing $\phl P$, the
context will contain $\dbr{\Psi}\br[u]{\pairQ yA}\pairQ xB$. That
still means that $\phl x$ must have type $B$ and $\phl y$ must have
type $A$ in $\phl P$, but also that $\pairQ yA$ has been received over
$\phl x$ (it is in $\phl x$'s queue) and we are intending to forward
it to endpoint $\phl u$. Moreover, $\Psi$ contains the types of
messages that have been previously received over $\phl x$.
The forwarders behave asynchronously. 
They can input arbitrarily many messages, which are enqueued at the arrival point, without blocking the possibility of producing an output from the same endpoint.
This behaviour is captured by the notion of queues of \emph{boxed}
messages, i.e.~messages that are in-transit.
\begin{smallequation*}
  \begin{array}{rcllllllll}
	\dbr {\Psi} & \coloncolonequals & \varnothing 
	& \mid\quad \br[u]{\Star}\dbr \Psi
	& \mid\quad \br[u]{\pairQ yA}\dbr \Psi
\end{array}
\end{smallequation*}
A queue element $\br[u]{\pairQ xA}$ expresses that some name $\phl x$
of type $\thl A$ has been received and will need to later be forwarded
to endpoint $\phl u$.
Similarly, $\br[u]{\Star}$ indicates that a request for closing a
session has been received and must be forwarded to 
$\phl u$.

The notation $\dbr{\Psi}$ is convenient as when needed we can refer to $\Psi$, the
set of all the elements that appear in $\dbr{\Psi}$ removing the
annotated brackets.
For example, for
	$\dbr{\Psi} =
	\br[v]{\pairQ{x}{A_1}}\br[u]{\pairQ{y}{A_2}}\br[w]{\Star}$, then
	$\Psi = \set{\pairQ{y}{A_2}, \pairQ{x}{A_1}, \Star}$.

The order of messages needing to be forwarded to \emph{independent}
endpoints is irrelevant.
Hence, we consider queue
$\dbr{\Psi_1}\br[x]{\ldots}\br[y]{\ldots}\dbr{\Psi_2}$ equivalent to
queue $ \dbr{\Psi_1}\br[y]{\ldots}\br[x]{\ldots}\dbr{\Psi_2}$
whenever $\phl x\not= \phl y$.
For a given endpoint $\phl x$ however the order of two messages
$\br[x]{\ldots}\br[x]{\ldots}$ is crucial and must be maintained
throughout the forwarding.
%
%
By attaching a queue to each endpoint we get a typing context.
\begin{smallequation*}
\begin{array}{rcllllllll}
  \Gamma & \coloncolonequals
  & \quad\varnothing
  & \quad\mid\quad \Gamma, \dbr \Psi \pairQ xB
  & \quad\mid\quad \Gamma, \dbr\Psi \pairQ x\cdot 
\end{array}
\end{smallequation*}
The element $\dbr{\Psi}\pairQ xB$ of a context $\Gamma$ indicates that
the messages in $\dbr{\Psi}$ have been received at endpoint $\phl x$.
The special case $\dbr{\Psi}\pairQ x\cdot$ is denoting the situation
when endpoint $\phl x$ no longer needs to be used for communication,
but still has a non-empty queue of messages to forward.

\paragraph{Judgements and rules.}
A judgement denoted by $\phl P \seq \Gamma$ types the forwarder processes $\phl P$ that connect the endpoints in $\Gamma$. 
The rules enforce the asynchronous forwarding behaviour by adding
elements to queues using 
rules for $\bot$ and $\parr$, which forces them to be later removed
from queues by the corresponding 
rules for $\one$ and $\tensor$.  
The rules are reported in Fig.~\ref{fig:mult-rules}.

\begin{figure}[t]
\begin{smallequation*}
	\begin{array}{c}
		\infer[\tsc{Ax}]{
			\fwd xy\seq\pairQ{x}{\dual a}, \pairQ ya
		}{}
		\qquad	
		\infer[\one]{ 
		\close x \seq \set[1\le i \le n]{\br[x]{\Star}\pairQ{u_i}{\cdot}}, \pairQ{x}{\one^\phl{\tilde u}}
		}{ }
		\\\\			
		\infer[\parr]{
			\phl{\recv xyP} 
			\seq 
			\Gamma, \dbr{\Psi}\pairQ x{A \parr^{\phl u} B} 
		}{
			\phl P 
			\seq 
			\Gamma, \dbr{\Psi}\br[u]{\pairQ yA}\pairQ xB
		}
		\qquad
		\infer[\bot]{
			\wait xP \seq \Gamma, \dbr{\Psi}\pairQ{x}{\bot^\phl{u}}
		}{
			\phl{P} \seq \Gamma, \dbr{\Psi}\br[u]{\Star}\pairQ{x}{\cdot}
		}
		\\\\
		\infer[\tensor]{
			\phl{\send xy{P}Q} 
			\seq 
			\Gamma,
			\br[x]{\pairQ{z}{\dual A}}\dbr{\Psi_u}\pairQ{u}{C},
			\dbr{\Psi_x}\pairQ x{A \tensor^\phl{u} B}
		}{
			\phl {P}
			\seq
			\pairQ{z}{\dual {A(y)}},\pairQ y{A(z )}
			& 
			\phl Q 
			\seq
			\Gamma, \dbr{\Psi_u}\pairQ{u}{C}, \dbr{\Psi_x}\pairQ xB
		}
	\end{array}
\end{smallequation*} 
\caption{Forwarder multiplicative rules}
\label{fig:mult-rules}
\end{figure}

Rule \tsc{Ax} is identical to the one of CP.
Rules $\one$ and $\bot$ forward a request to close a session. Rule $\bot$ receives the request on endpoint $\phl x$ and enqueues it as $\br[u]{\Star}$ if it needs to forward it to $\phl u$. Note that in the premiss of $\bot$ the endpoint is terminated pending the remaining messages in the corresponding queue being dispatched. Eventually all endpoints but one will be terminated in the same manner. 
Rule $\one$ will then be applicable. Note that the behaviour of $\wait xP$ and $\close x$ work as gathering, 
several terminated endpoints connect to the last active endpoint typed with a $\one$.
%
Rules $\tensor$ and $\parr$ forward a message. Rule $\parr$ receives the message $\pairQ y A$ and enqueues it as $\br[u]{\pairQ y A}$ to be forwarded to endpoint $\phl u$. 
Dually, rule $\tensor$ applied to a $\tensor^{\phl u}$ sends the message at the top of the queue of endpoint $\phl u$ if it has the dual type.
Note that this idea can be generalised to a gathering behaviour where several messages are sent at the same time. Messages will be picked from queues belonging to distinct endpoints, which would require us to annotate the tensor with a list of endpoints. As a consequence, the left premiss of $\tensor$ rule would be a new forwarder consisting of the gathered messages. For the sake of simplicity we discuss this generalisation only in the appendix.

Note how annotations put constraints on how the proof is constructed, e.g., annotating an $\pairQ x{A\parr B}$ with $\phl u$ ensures us that the proof will contain a $\tensor$-rule application on endpoint $\phl u$ at a later point.

\begin{example}\label{ex:criss-cross}
	$\phl P \colonequals \recv xu{\recv yv{\send y{u'}{\fwd u{u'}}{\send x{v'}{\fwd {v'}v}{\wait x{\close y}}}}}$
	is one of the forwarders that can prove the compatibility of the types involved in the criss-cross protocol (in \S~\ref{sec:preview}), 
	as illustrated by the derivation below in forwarder logic.
	
	\begin{smallequation*}
	\infer[\parr]{
		\phl P
		\seq
		\pairQ x{\tname^\perp\parr \tcost ^\perp\tensor \bot},
		\pairQ y{\tcost\parr  \tname\tensor \one}
		}{
		\infer[\parr]{
			\recv yv{\send y{u'}{F_1}{\send x{v'}{F_2}{F_3}}}
			\seq
			\br[y]{\pairQ{u}{\tname^\perp}}\pairQ x{ \tcost ^\perp\tensor \bot},
			\pairQ y{\tcost\parr  \tname\tensor \one}
			}{
			\infer[\tensor]{
				\send y{u'}{F_1}{\send x{v'}{F_2}{F_3}}
				\seq
				\br[y]{\pairQ{u}{\tname^\perp}}\pairQ x{ \tcost ^\perp\tensor \bot},
				\br[x]{\pairQ{v}{\tcost}}\pairQ y{\tname\tensor \one}
				}{
				\infer[\tsc{Ax}]{
					\seq
					\pairQ{u}{\tname^\perp}, \pairQ{u'}{\tname}
					}{
					\phl{F_1} \colonequals \fwd u{u'}
					}
					&
					\infer[\tensor]{
						\send x{v'}{F_2}{F_3}
						\seq 
						\pairQ x{ \tcost ^\perp\tensor \bot},
						\br[x]{\pairQ{v}{\tcost}}\pairQ y{\one}
						}{
						\infer[\tsc{Ax}]{
							\seq
							\pairQ{v'}{\tcost ^\perp}, \pairQ{v}{\tcost}
							}{
							\phl{F_2} \colonequals \fwd {v'}v
							}
						&
						\infer[\bot]{
							\phl{F_3}
							\seq 
							\pairQ x{\bot},
							\pairQ y{\one}
							}{
							\infer[\one]{
								\close y
								\seq 
								\br[y]{\Star}\pairQ x{\cdot},
								\pairQ y{\one}
								}{
								\phl{F_3} \colonequals \wait x{\close y}
								}
							}
						}
				}
			}
		}
	\end{smallequation*}
\end{example}

We conclude this section by stating that every forwarder is also a CP
process, the embedding $\enc{\cdot}$ being extended to contexts as:
\begin{smallequation*}
	\begin{array}{l@{\qquad\qquad}l@{\qquad}l}
		\enc{\dbr{\Psi}\pairQ{x}{B}, \Gamma} = \enc{\dbr{\Psi}}, \pairQ{x}{\enc B}, \enc{\Gamma}
		&
		\enc{\dbr{\Psi}\pairQ{x}{\cdot}, \Gamma} = \enc{\dbr{\Psi}}, \enc{\Gamma}
		\\[1ex]
		\enc{\br[u]{\pairQ{y}{A}}\dbr{\Psi}}  = \pairQ{y}{A} , \enc{\dbr{\Psi}}
		&
		\enc{\br[u]{\Symbol}\dbr{\Psi}}  = \enc{\br[u]{\Star}\dbr{\Psi}} = \enc{\dbr{\Psi}}
	\end{array}
\end{smallequation*}

\begin{proposition}\label{prop:embed}
	Any forwarder is typable in CP, i.e., if $\phl P \seq \Gamma$, then  $\phl P \cll \enc{\Gamma}$.
\end{proposition}

%
%

\section{Semantics of Asynchronous Forwarders}\label{sec:semantics}
Our next task is to lay the groundwork that will eventually allow us
to establish a semantics for this system via a cut elimination
procedure.
%
%
Adding label to queues and connectives brings along some notational
burden, but it is instrumental to complete the cut-elimination proof.
However, this notational overhead comes at a price, since the labels
need to be meticulously maintained: the cut-elimination procedure does
not only simplify the formulas and derivations involved in the cut but
also rewrites the labels used within a sequent.

Elements of a context are formulas which ($i$) feature annotated
connectives and ($ii$) are equipped with queues of messages to be
forwarded; these two new features need to appear in the standard cut
rule of linear logic, and, as a consequence, pose new challenges to
the cut-elimination procedure.
A cut rule must cut two elements of a context of the form $\dbr{\Psi_1}\pairQ x A$ and $\dbr{\Psi_2}\pairQ y{\dual A}$.
As shown by Proposition~\ref{prop:embed}, in a standard linear logic cut rule, the elements of $\Psi_1$ and $\Psi_2$ would freely occur in the conclusion context as independent formulas. 
However, in forwarder logic, they are attached to $\phl x$ and $\phl y$, respectively, and they are the type of messages that must be forwarded. 
Hence, after cutting $\pairQ x A$ and $\pairQ y{\dual A}$ we must attach $\Psi_1$ and $\Psi_2$ to other endpoints in the context of the conclusion. 
This operation is called \emph{distribution}.

Moreover, since queue elements in the rest of the context are also annotated with endpoints, we must handle those referring  to $\phl x$ and $\phl y$, which are doomed to disappear after the cut. 
The \emph{substitution} operation does this guided by the annotations on the cut formulas.
%


Both operations \emph{substitution} and  \emph{distribution} work on a two dimensional depiction of the cut judgements $\Gamma_1, \dbr{\Psi_1}\pairQ xA$ and $\Gamma_2, \dbr{\Psi_2}\pairQ y{\dual A}$ as follows:
\begin{smallequation}\label{eq:cut-pair}
\left\{
\begin{array}{l@{\quad\rhd\quad}l} 
	\Gamma_1 & \dbr{\Psi_1}\pairQ xA\\
	\Gamma_2 & \dbr{\Psi_2}\pairQ y{\dual A}
\end{array}
\right\}
\end{smallequation}
We use the $\rhd$ to single out the cut-formulas and make the rules
that we introduce below more readable. $\rhd$ should not be confused
with derivability in intuitionistic logic.  Moreover, when we need to
single out an operator, for instance to substitute a different name
for its annotation, we use the notation $B\ctx{\tensor^\phl{u}}$, this
indicates the first (leftmost) occurrence of the symbol
$\tensor^{\phl u}$ within proposition $B$.

\begin{example}
	The proposition $B \colonequals(a \tensor b) \tensor^\phl{u} (c \parr d) \tensor^\phl{u} e$ is a well-formed annotated type. 
	We would write $B\ctx{\tensor^\phl{u}}$ to point to the leftmost occurrence of $\tensor^\phl{u}$.
	Consequently $B\ctx{\tensor^\phl{x}}$ would denote the formula $(a \with b) \tensor^\phl{x} (c \parr d) \tensor^\phl{u} e$.
	On the other hand, $B[\phl x/\phl u]$ would indicate a global substitution of $\phl x$ for $\phl u$, namely the formula
	$(a \tensor b) \tensor^\phl{x} (c \parr d) \tensor^\phl{x} e$.
\end{example}

\mypar{Distribution.}
Given a pair of contexts depicted as in~\eqref{eq:cut-pair},
%
we proceed by distributing each element of the queues $\dbr{\Psi_1}$ and $\dbr{\Psi_2}$ to queues in $\Gamma_2$ and $\Gamma_1$ respectively.
Indeed, elements in $\dbr{\Psi_1}$ ($\dbr{\Psi_2}$) have been received at endpoint $\phl x$ ($\phl y$), but when executing the cut $\phl x$ and $\phl y$ disappear and only the other endpoints in $\Gamma_1$ and $\Gamma_2$ remain.
Messages in $\Psi_1$ ($\Psi_2$) need now to come from an available endpoint from $\Gamma_2$ ($\Gamma_1$).
We always distribute by picking the top element (in our notation
below, a box labelled with $d$) of one of the queues and since the cut
rule is symmetric, we pick from $\Psi_1$.

We start with equipping each queue on the left with a $\bullet$ marker
that ensures that the messages from successive applications of the $\mathsf{distr}$ rule preserve their respective orders.
Then, assuming that part of the queues have already been distributed, 
the following is the $\mathsf{distr}$ rewriting that allows us to distribute the top element of the queue in front of endpoint $\phl x$ to endpoint $\phl c$ in $\Gamma_2$
%
\begin{smallalign*}
	\left\{
	\begin{array}{l@{\quad\rhd\quad}l}
		\Gamma_1, \dbr{\Psi_{2d}}\bullet\dbr{\Psi_d}\pairQ{d}{D\ctx{\tensor^\phl{x}}}  & \br[d]{\pairQ bB}\dbr{\Psi_1}\pairQ{x}{A}
		\\
		\Gamma_2, \dbr{\Psi_{1c}}\bullet\dbr{\Psi_c}\pairQ{c}{C}& \dbr{\Psi_2}\pairQ{y}{\dual A}
	\end{array}
	\right\}
	\xrightarrow{}_{\mathsf{distr}} 
	\\
	\left\{
	\begin{array}{l@{\quad\rhd\quad}l}
		\Gamma_1, \dbr{\Psi_{2d}}\bullet\dbr{\Psi_d}\pairQ{d}{D\ctx{\mathbf{\tensor}^\phl{c}}} & \dbr{\Psi_1}\pairQ{x}{A}\\
		\Gamma_2, \dbr{\Psi_{1c}}\br[d]{\pairQ bB}\bullet\dbr{\Psi_c}\pairQ{c}{C} & \dbr{\Psi_2}\pairQ{y}{\dual A}
	\end{array}
	\right\}       
\end{smallalign*}
%
%


An application  of this rule yields again two sequents.  
We move $\br[d]{\pairQ bB}$ from the queue in front of $\pairQ xA$ to endpoint $\phl c$ and the left hand side of $\rhd$ in the top sequent is updated to reflect that $\phl d$ has one fewer communication link with $\phl x$ and communicates with  channel $\phl {c}$ instead, i.e., $\pairQ{d}{D\ctx{\tensor^\phl{x}}}$ becomes $\pairQ{d}{D\ctx{\mathbf{\tensor}^\phl{c}}}$.
Eventually, when both queues $\dbr{\Psi_1}$ and $\dbr{\Psi_2}$ have been distributed, the obsolete $\bullet$ markers are removed and the process enters the next phase, substitution. 

Note that the $\mathsf{distr}$ rule is not deterministic, in that the rule does not  uniquely determine, to which endpoint the element  $\br[d]{\pairQ{b}{B}}$ is distributed to.  
%
%
Hence, each application of $\mathsf{distr}$ will determine a particular cut-rule.

\mypar{Substitution.}
After distribution, the cut-formulas are free from their respective
queues which have been placed over endpoints in the rest of the
contexts. The substitution operation is now going to take care of all
those occurrences of $\phl x$ and $\phl y$ in the contexts that still
must disappear by exploiting the annotations on the cut-formula.
We start from a configuration of the following form:
\begin{smallequation*}
\left\{
\begin{array}{l@{\quad\rhd\quad}l} 
	\Gamma_1 & \pairQ xA\\
	\Gamma_2 & \pairQ y{\dual A}
\end{array}
\right\}
\end{smallequation*}

The substitution operation is defined inductively following the respective structures of  $\pairQ x{A}$ and $\pairQ y{\dual A}$.  We distinguish three cases, the $\tensor/\parr$ case, the atomic case, and the unit case.

\paragraph{Substitution for $\tensor$ and $\parr$.}
Assume that the cut formulas are of the form
$\pairQ{x}{A \tensor^{\phl{b}} D} $ and
$ \pairQ{y}{\dual A \parr^{\phl{c}} \dual D}$.
This enforces that, in the bottom left context ($\Gamma_2$), the endpoint $\phl c$ must always be typed by a proposition that contains a $\tensor^\phl{y}$. The $\phl y$ will need to be substituted by $\tensor^\phl{b}$.
In the top left context ($\Gamma_1$) however, there can be two
scenarios: either the endpoint $\phl b$ has an element such as
$\br[\phl{x}] {\pairQ{a}{\dual A}}$ in its queue, i.e.~typed with the
dual of $A$ and that needs to be forwarded to $\phl x$, or it has a
$\parr^\phl{x}$ within its type. In both cases, $\phl x$ needs to be
substituted by $\phl c$.

In the first case, 
it gives the following rewriting
\begin{smallalign*}
	\left\{
\begin{array}{l@{\quad\rhd\quad}l}
	\Gamma_1, \br[x] {\pairQ{a}{\dual A}}\dbr{\Psi_b}\pairQ{b}{B} 
	& 
	\pairQ{x}{A \tensor^{\phl{b}} D} 
	\\
	\Gamma_2, \dbr{\Psi_c}\pairQ{c}{C\ctx{\tensor^\phl{y}}} 
	&
	\pairQ{y}{\dual A \parr^{\phl{c}} \dual D} 
\end{array}
\right\}
\xrightarrow{}_{\mathsf{subst}} 
\\
\left\{
\begin{array}{l@{\quad\rhd\quad}l}
	\Gamma_1, \br[c]{\pairQ{a}\dual{A}}\dbr{\Psi_b}\pairQ{b}{B} 
	& \pairQ{x}{D} \\
	\Gamma_2, \dbr{\Psi_c}\pairQ{c}{C\ctx{\tensor^\phl{b}}}
	&
	\phl y: \thl {D^\perp} 
\end{array}
\right\} 
\end{smallalign*}

In the second case, it gives a similar rewriting  

\begin{smallalign*}
	\left\{
\begin{array}{l@{\quad\rhd\quad}l}
	\Gamma_1, \dbr{\Psi_b}\pairQ{b}{B\ctx{\parr^\phl{ x}}}  
	& \pairQ{x}{A \tensor^{\phl{ b}} D} \\
	\Gamma_2, \dbr{\Psi_c}\pairQ{c}{C\ctx{\tensor^\phl{y}}}   &
	\pairQ{y}{\dual A \parr^{\phl{ c}} \dual D} 
\end{array}
\right\}
\xrightarrow{}_{\mathsf{subst}}
\\
\left\{
\begin{array}{l@{\quad\rhd\quad}l}
	\Gamma_1, \dbr{\Psi_b}\pairQ{b}{B\ctx{\parr^\phl{c}}}  & \pairQ{x}{D} \\
	\Gamma_2,  \dbr{\Psi_c}\pairQ{c}{C\ctx{\tensor^\phl{b}}} &
	\phl y: \thl {D^\perp}
\end{array}
\right\}
\end{smallalign*}


\paragraph{Substitution for atoms.}
Substituting atoms is straightforward and completes the substitutions
sequence, by producing the standard judgement.

\begin{smallequation*}
	\left\{
\begin{array}{l@{\quad\rhd\quad}l}
	\Gamma_1 & \pairQ{x}{a} \\
	\Gamma_2 & \pairQ{y}{\dual a}
\end{array}
\right\}
\xrightarrow{}_{\mathsf{subst}}
	\Gamma_1, \Gamma_2
\end{smallequation*}

\paragraph{Substitution for units.}
Substitution for units resemble the multiplicative connective but with
the added feature of gathering (for a treatment of the multiplicative
connective that includes gathering, see the appendix), but it
terminates the rewriting sequence similarly to the atoms.
Assume the cut formulas are $\pairQ{x}{\one^\phl{\tilde a\tilde b}}$ and $\pairQ{y}{\bot^\phl{c}} $.
This means that the top context ($\Gamma_1$) can potentially contain some terminated endpoints (here, the $a_i$'s), equipped with a  queue of the form $\dbr{\Psi_{a_i}}\br[x]{\Star}$ and some other channels (here, the $b_j$'s) whose type includes a $\bot^\phl{x}$. These $\phl x$ will need to be substituted by $\phl c$.
It also means that the endpoint $\phl c$'s type,  in the bottom context ($\Gamma_2$), embeds a unit $\one$ labelled with a set of endpoints which includes $\phl y$. This $\phl y$ will need to be replaced by the list on endpoints $\phl{\tilde a\tilde b}$.
This gives us the following rewriting

\begin{smallalign*}
	\left\{
\begin{array}{l@{\quad\rhd\quad}l}
	\Gamma_1,\set[i]{ \dbr{\Psi_{a_i}}\br[x]{\Star}\pairQ{a_i}{\cdot}}, \set[j]{\dbr{\Psi_{b_j}}\pairQ{b_j}{B_j\ctx{\bot^\phl{x}}}}  & 
	\pairQ{x}{\one^\phl{\tilde a\tilde b}}\\
	\Gamma_2,\dbr{\Psi_c}\pairQ{c}{C\ctx{\one^\phl{y\tilde{u}}}} & \pairQ{y}{\bot^\phl{c}} 
\end{array}
\right\}
\xrightarrow{}_{\mathsf{subst}}
\\
\Gamma_1, \set[i]{\dbr{\Psi_{a_i}}\br[c]{\ast}\pairQ{a_i}{\cdot}}, \set[j]{\dbr{\Psi_{b_j}}\pairQ{b_j}{B_j\ctx{\bot^\phl{c}}}},
\Gamma_2,\dbr{\Psi_c}\pairQ{c}{C\ctx{\one^\phl{\tilde a\tilde b\tilde{u}}}}
\end{smallalign*}

In summary, the definition of the cut-rule is a bit more complicated than in classical linear logic: What used to be a simple identification of the cut-formula in the left sequent and its dual on the right, has become a sequence of somewhat cumbersome distribution and substitution steps. Hence, we will show how to adapt the proof accordingly in the rest of this section.

\mypar{Reductive semantics.}
The \cut is defined in terms of distribution and substitution
(applying these rewritings until they come to quiescence) and
represents the interaction between two active processes $\phl P$ and
$\phl Q$.
Recall that due to the non-determinism of the $\mathsf{distr}$-rule, each distribution sequence will determine a different conclusion $\Gamma$.
%

\begin{smallequation*}
		\infer[\cut]
	{\phl {\res{xy} (P\pp Q)}\seq^\cut \Gamma}
	{
			\phl P\seq \Gamma_1, \dbr{\Psi_1}\pairQ xA
& 
		\phl Q\seq\Gamma_2, \dbr{\Psi_2}\pairQ y{A^\perp}
&
		\big\{\Gamma_1\rhd\dbr{\Psi_1}\pairQ xA,
		\Gamma_2\rhd\dbr{\Psi_2}\pairQ y{\dual
			A}\big\}\rightarrow^*_{\mathsf{distr}}\rightarrow^*_{\mathsf{subst}}\Gamma
} 
\end{smallequation*}

We denote by $\phl P \seq^\cut \Gamma$ the fact that a derivation can be constructed using the rules previously introduced for forwarders as well as the additional $\cut$ rule.
The rank of a $\cut$ is defined as the size (number of connectives and units) of $A$. 
By extension the cut-rank of process $\phl P$ such that $\phl P\seq^\cut\Gamma$, denoted as $\mathtt{rank}(\phl P)$, is the maximum of the ranks of $\cut$ rules occurring in the derivation of this judgement. 


As one could hope in the proposition-as-type methodology, semantics of forwarders is obtained through cut-reductions.

\begin{theorem} [Admissibility of \tsc{Cut}]
  Let $\phl P\seq\Gamma_1, \dbr{\Psi_1}\pairQ xA$ and
  $\phl Q\seq\Gamma_2, \dbr{\Psi_2}\pairQ y{A^\perp}$.  
  Then, for
  any $\Gamma$ such that
  $\big\{\Gamma_1\rhd\dbr{\Psi_1}\pairQ xA,\ \Gamma_2\rhd\dbr{\Psi_2}\pairQ y{\dual
    A}\big\}\longrightarrow^*_{\mathsf{distr}}\longrightarrow^*_{\mathsf{subst}}\Gamma$
  there exists $\phl R \seq \Gamma$,
 more precisely, we can define a sequence of reductions $\phl{\res{xy}(P \pp Q)} \reducesto^*_\beta \phl R$ that preserves typing.
\end{theorem}

\begin{figure}[t]
\begin{smallequation*}
		\begin{array}{l@{\qquad}l@{\quad\reducesto_\beta\quad}l}
		(B_1) &
		\phl{\res{xy}(\fwd zx \pp Q)}
		&
		\phl{Q[z/y]}
		\\
		(B_2) &
		\phl{\res {xy}(\close x \pp \wait {y}{Q}  )} 
		&
		\phl{Q^\mathsf{y \leadsto \tilde u}} 
		\\
		(C_1) &
		\phl{\res{xy}(P \pp \wait zQ)}
		&
		\wait{z}{ \phl{\res{xy}(P \pp Q)}}
		\\
		(C_2)
		&
		\phl {\res{xy} (P\pp \recv uvQ)}
		&
		\recv uv{\res {xy}(P\pp Q)} 
		\\
		(C_3) &
		\phl{\res {xy}(P\pp (\send z{v}{Q}{R}))}
		&
		\send zvQ{\res{xy}(P \pp R)}
		\\
		(K) &
		\phl{\res{xy}(\send xa{P}{Q} \pp \recv yc{R}) }
		&
		\phl {\res{xy} (Q\pp \res{a\rhd c} (P\pp R))}
	\end{array}
\end{smallequation*}
	\caption{Cut-reductions for the multiplicative fragment}
	\label{fig:cut-red}
\end{figure}

	We will describe the method of the proof, for the details see the appendix.
	It proceeds by lexicographic induction on the structures of $A$, $\phl P$ and $\phl Q$.
	That is, the induction hypothesis may be applied whenever 
	($i$) the rank of the cut gets smaller,
	or ($ii$) the rank stays the same and the cut is applied to at least one smaller process while the other stays the same.

As usual, we can distinguish (B)ase cases, (K)ey cases and (C)ommutative cases, see Figure~\ref{fig:cut-red}. 
In the base cases ($B_1$) and ($B_2$), we reach the end of the reduction sequence and the cut disappears all together.
In ($B_1$) we replace the cut with a simple uniform substitution; 
in ($B_2$)  on the other hand, we require a non-uniform substitution reproducing the $\mathsf{distr}$ rewriting happening in the original cut.
In the commutative cases ($C_1$), ($C_2$) and ($C_3$), the cut is reduced to a cut on $\phl P$ (that remains the same) and on a subprocess of $\phl Q$, to which the induction hypothesis can easily be applied.
Identical reductions are of course available symmetrically inverting the roles of $\phl P$ and $\phl Q$.
In the key case, the cut is rewritten into a cut of smaller rank on process $\phl Q$ and $\phl S = \phl{\res{a \rhd c}(P \pp R)}$. 
This process $\phl S$ is obtained by Lemma~\ref{lem:cuttwo-adm-simp} (see appendix).

The last piece of the puzzle that we need to complete the cut-elimination proof is a property of the distribution/substitution rewriting. 
%
%
To ensure that the type of  $\phl {\res{xy} (Q\pp \res{a\rhd c} (P\pp R))}$ is indeed the same as the one of the original process $\phl{\res{xy}(\send xa{P}{Q} \pp \recv yc{R}) }$, we rely on the fact that, if
\begin{smallequation*}
	\left\{
\begin{array}{l@{\quad\rhd\quad}l}
	\Gamma_1,
	\br[x]{\pairQ{d}{\dual A}}\dbr{\Psi_u}\pairQ{u}{C}  & 
	\dbr{\Psi_1}\pairQ x{A \tensor^\phl{u} B}\\
	\Gamma_2, \dbr{\Psi_v}\pairQ{v}{D\ctx{\tensor^\phl{y}}} &  \dbr{\Psi_2}\pairQ y{\dual A \parr^\phl{v} \dual B} 
\end{array}
\right\}
\xrightarrow{}_{\mathsf{distr}}^*\xrightarrow{}_{\mathsf{subst}}^*
\Gamma
\end{smallequation*}
then there exists a distribution-substitution rewriting sequence such that 
\begin{smallequation*}
	\left\{
\begin{array}{l@{\quad\rhd\quad}l}
	\Gamma_1, \dbr{\Psi_u}\pairQ{u}{C} &
	\dbr{\Psi_1}\pairQ x{B}\\
	\Gamma_2, \dbr{\Psi_v}\pairQ{v}{D\ctx{\tensor^\phl{y}}} &  \dbr{\Psi_2}\br[v]{\pairQ{d}{\dual A}}\pairQ y{\dual B}
\end{array}
\right\}
\xrightarrow{}_{\mathsf{distr}}^*
\xrightarrow{}_{\mathsf{subst}}^*
\Gamma
\end{smallequation*}
This can be established by induction over the length of the rewriting sequence.

%
%

\section{Asynchronous Forwarders Generalise Coherence}\label{sec:mcut}
So far, we have focused on asynchronous forwarders syntax and
semantics. 
As already explained intuitively in Section~\ref{sec:preview},
one of the main contributions of this work is to use
forwarders as a medium among communicating processes 
(more precisely those typable in the system CP presented in Section~\ref{sec:prelim}). 
%



\mypar{Multiparty Process Composition.}  We start by
focusing on the {\em structural} rule that can be added to CP, namely the $\cut$, 
as seen in Section~\ref{sec:prelim}. 
After that, we will introduce an alternative (more general) rule
that makes use of asynchronous forwarders. 
Rule \tsc{Cut} corresponds to parallel
composition of processes. The implicit side condition that this rule
uses is \emph{duality}, i.e., we can compose two
processes if endpoints $\phl x$ and $\phl y$ have a dual type. Carbone
et al.~\cite{CLMSW16} generalise the concept of duality 
to that of {\em coherence}. Coherence,
denoted by $\coherence$, generalises duality to many endpoints,
allowing for a cut rule that composes many processes in parallel 
\begin{smallequation*}
  \infer[\tsc{MCut}]
  {\phl{\res{\tilde x:G} (R_1 \pp \ldots \pp R_n)} \cll \set[i\le n]{\Sigma_i}}
  {
    \set[i\le n]{\phl {R_i} \cll \thl\Sigma_i, \pairQ{x_i}{A_i}}
    &
    \phl G \coherence \thl\set[i\le n]{\pairQ{x_i}{A_i}}}
\end{smallequation*}
The judgement $\phl G \coherence \set[i\le n]{\pairQ{x_i}{A_i}}$ intuitively
says that the $\pairQ{x_i}{A_i}$'s are compatible and the execution of the
$\phl{R_i}$ will proceed without any error (no deadlock, no type mismatch in
messages). Such a result is formalised by an \tsc{MCut} elimination
theorem analogous to the one of CP. We leave $\phl G$ abstract here: it is a proof term and it
corresponds to a global type (see~\cite{CLMSW16}).

Our goal here is to replace the notion of coherence with an asynchronous
forwarders $\phl Q$, hoping for a rule resembling the following
\begin{smallequation*}
  \infer[\tsc{MCutF}]
  {\phl{\res{\tilde x:Q}(R_1 \pp \ldots \pp R_n)} \cll \set[i\le n]{\Sigma_i}}
  {
    \set[i\le n]{\phl {R_i} \cll \thl\Sigma_i, \pairQ{x_i}{A_i}}
    &
    \phl Q \seq \thl\set[i\le n]{\pairQ{x_i}\dual{A_i}}
  }
\end{smallequation*}
Asynchronous forwarders are more general than coherence: every
coherence proof can be transformed into an \emph{arbiter} process~\cite{CLMSW16}, which is indeed a forwarder,
while there are judgements that are not coherent but are provable in
our forwarder logic (see Example~\ref{ex:criss-cross}). 
In the rule \tsc{MCutF},
the role of a forwarder (replacing coherence) is to be a
middleware that decides whom to forward messages to. This means that
when a process $\phl{R_i}$ sends a message to the middleware, the 
message must be stored by the forwarder, who will later
forward it to the right receiver.

Since our goal is to show that \tsc{MCutF} is admissible (and hence we
can eliminate it from any correct proof), we extend such rule to
account for messages in transit that are temporarily held by the
forwarder. In order to do so, we use the forwarders queues and some
extra premises and define $\tsc{MCutQ}$ as:
\begin{smallequation*}
  \infer[]{
  	\phl{\res{\tilde x: Q \br{\tilde y \lhd P_1, \ldots, P_m}} (R_1 \pp \ldots \pp R_n)} \cll \set[j\le m]{\Delta_j}, \set[i\le n]{\Sigma_i}
  }{
  	\set[j\le m]{\phl {P_j} \cll \thl\Delta_j, \pairQ{y_j}{A_j}}
    &
    \set[i\le n]{\phl {R_i} \cll \thl\Sigma_i, \pairQ{x_i}{B_i}}
    &
    \phl Q \seq \set[i\le n]{\dbr{\Psi_i}\pairQ{x_i}\dual{B_i}}, \set[n < i \le p]{ \dbr{\Psi_i}\pairQ{x_i}{\cdot}}
  }
\end{smallequation*}
We have three types of process terms:
$\phl{P_j}$'s, $\phl{R_i}$'s and $\phl{Q}$. Processes $\phl{R_i}$'s
are the processes that we are composing, implementing a multiparty
session. $\phl Q$ is the forwarder whose role is to certify
compatibility and to determine, at run time, who talks to
whom. Finally, processes $\phl{P_i}$'s must be linked to messages in the
forwarder queue. 
Such processes are there because of the way $\tensor$
and $\parr$ work in linear logic. This will become clearer when we
look at the reduction steps that lead to cut admissibility. 
This imposes a side condition on the rule, namely that 
\begin{smallequation*}
	\bigcup_{i\le p}{\Psi_i} \setminus
\set{\Star} = \set[j\le 
m]{\pairQ{y_j}{\dual{A_j}}}
\end{smallequation*}
Note that we need to introduce a new syntax for this new structural rule: 
in $\phl{\res{\tilde x: Q \br{\tilde y \lhd P_1, \ldots, P_m}} (R_1 \pp \ldots \pp R_n)}$, 
the list $\phl{P_1, \ldots, P_m}$ denotes those messages (processes) in transit
that are going to form a new session after the communication has taken place.
In the remainder we (slightly abusively) abbreviate both  $\phl{\set{P_1, \ldots, P_m}}$ and $\phl{(R_1 \pp \ldots \pp R_n)}$ as $\phl{\tilde P}$ and $\phl{\tilde R}$ respectively.

\mypar{Semantics and \tsc{MCutF}-admissibility.} In the previous
paragraph, we have informally argued that forwarders generalise the
notion of coherence as a notion of compatibility for composing
processes typable in classical linear logic. In order to do that
formally, we show that \tsc{MCutF} is admissible, yielding a semantics
for our extended CP (with \tsc{MCutF}) in a
proposition-as-types fashion.

We proceed by looking at all cases that involve the multiplicative
fragment (see appendix for the full set of rules).
In the sequel, we use the following abbreviations,
$\Gamma = \set[i\le n]{\dbr{\Psi_i}\pairQ{x_i}\dual{B_i}}, \set[n < i
\le p]{ \dbr{\Psi_i}\pairQ{x_i}{\cdot}}$ and
$\Gamma{-k}= \Gamma \setminus
\set{\dbr{\Psi_k}\pairQ{x_k}\dual{B_k}}$.
We also omit (indicated as ``$\ldots$'') the premises of the
$\tsc{MCutQ}$ that do not play a role in the reduction at hand, and
assume that they are always the same as above, that is,
$\set[j\le m]{\phl {P_j} \cll \thl\Delta_j, \pairQ{y_j}{A_j}}$ and
$\set[i\le n]{\phl {R_i} \cll \thl\Sigma_i, \pairQ{x_i}{B_i}}$.

\noindent {\it Send Message ($\tensor$).} This is the case when a
process intends to send a message, which corresponds to a
$\tensor$ rule. As a consequence, the forwarder has to be ready to
receive the message (to then forward it later):
\begin{smallequation*}
  \infer[\tsc{MCutQ}]{
  	\phl{\res{x\tilde x:\recv xyQ\br{{\tilde y} \lhd\tilde P}}(\send{x}{y}{P}{R} \pp \tilde R)} \cll \Delta, \Sigma, \set[j\le m]{\Delta_j}, \set[i\le n]{\Sigma_i}
  }
  {
    \infer[\tensor]
    {
      \phl{\send{x}{y}{P}{R}}\cll \Delta,\Sigma, \pairQ x{A\tensor B}
    }
    {
      \phl P\cll \Delta,\pairQ yA
      &
      \phl R\cll \Sigma,\pairQ xB
    }
    & \quad\ldots\quad
    &
    \infer[\parr]
    {
      \phl{\recv xyQ} \seq 
      \dbr{\Psi}\pairQ x{\dual A\parr^{\phl{x_k}}\dual B}, 
      \Gamma
    }
    {
      \phl Q\seq \dbr{\Psi}\br[x_k]{\pairQ y{\dual A}}\pairQ x{\dual B}, 
      \Gamma
    }
  }
\end{smallequation*}
The process on the left is ready to send the message to the forwarder.
By inspecting the forwarder, it is clear that the message will have to
be forwarded to endpoint $\phl{x_k}$, at a later stage. Observe that
the nature of $\tensor$ forces us to deal with the process $\phl P$:
the idea is that when the forwarder will finalise the communication
(by sending to a process $\phl{R'}$ owning endpoint $\phl{x_k}$)
process $\phl P$ will be composed with $\phl {R'}$. For now, we obtain
the reductum:
\begin{smallequation*}
  \infer[\tsc{MCutQ}]{
  	\phl{\res{x\tilde x:Q\br{{y,\tilde y} \lhd P, \tilde P}}(R \pp \tilde R)} \cll \Delta, \Sigma, \set[j\le m]{\Delta_j}, \set[i\le n]{\Sigma_i}
  }
  {
    	\phl P\cll\Delta,\pairQ yA
    	&
    	\phl R\cll\Sigma,\pairQ xB
    & \quad\ldots\quad
    &
    \phl Q\seq 
    \dbr{\Psi}\br[x_k]{\pairQ y\dual{A}}\pairQ x\dual{B}, 
    \Gamma
  }
\end{smallequation*}

\noindent {\it Receive Message ($\parr$).} At a later point,
the forwarder will be able to complete the forwarding operation by
connecting with a process ready to receive ($\parr$ rule):
\begin{smallequation*}
  \infer[\tsc{}]
  {\phl{\res{x\tilde x:\send xy{S}Q\br{z,\tilde y \lhd P,\tilde P}}(\recv{x}{y}R \pp \tilde R)}
  	\cll 
  	\Delta, \Sigma, \set[j\le m]{\Delta_j}, \set[i\le n]{\Sigma_i}
  }
  {
    \begin{array}{c}
      \phl{P} \cll \Delta, \pairQ{z}\dual{A}
  \hspace*{2cm}
  \ldots
  \hspace*{2cm}
    \\ 
    \infer[\parr]
    {
    	\phl{\recv{x}{y}{R}}\cll \Sigma, \pairQ x{A\parr B}
    }
    {
    	\phl{R}\cll \Sigma, \pairQ yA, \pairQ x{B}
    }
	\quad
    \infer[\tensor]{
    	\phl{\send xy{S}Q} 
    	\seq 
    	\dbr{\Psi_x}\pairQ x{\dual A \tensor^{\phl{x_k}} \dual B}, \br[x]{\pairQ{z}{A}}\dbr{\Psi_k}\pairQ{x_k}\dual{B_k},
    	\Gamma-k
    }{
    	\phl S
    	\seq
    	\pairQ{z}{A}, \pairQ y{\dual A}
    	&
    	\phl Q
    	\seq
    	\dbr{\Psi_x}\pairQ x{\dual B}, 
    	\Gamma
    }
    \end{array}
  }
\end{smallequation*}
Key ingredients are process $\phl P$ with endpoint $\phl z$
of type $\dual A$, endpoint $\phl{x_k}$ in the forwarder with a
boxed endpoint $\phl z$ with type $A$, and process
$\phl{\recv{x}{y}{R}}$ ready to receive. 

After reduction, we obtain the following:
\begin{smallequation*}
  \infer[\tsc{MCutQ}]
  {\phl{\res{x\tilde x:Q\br{\tilde y \lhd \tilde P}}(\res{yz:S}(R\pp P) \pp \tilde R)} \cll 
  \Delta, \Sigma, \set[j\le m]{\Delta_j}, \set[i\le n]{\Sigma_i}
	}
  {
      \phl{\res{yz:S}(R\pp P)}\cll\Sigma, \Delta, \pairQ xB
    %
    &\quad\ldots\quad
    &
    \phl Q
    \seq
    \dbr{\Psi_x}\pairQ x{\dual B}, 
    \Gamma
  }
\end{smallequation*}
Where the left premiss is obtained as follows:
\begin{smallequation*}
	    \infer[\tsc{MCutQ}]
	{
		\phl{\res{yz:S}(R\pp P)}\cll\Sigma, \Delta, \pairQ xB
	}
	{
		\phl R\cll \Sigma, \pairQ yA, \pairQ x{B}
		&
		\phl P\cll\Delta,\pairQ z{\dual A}
		&
		\phl S\seq\pairQ zA,\pairQ y{\dual A}
	}
\end{smallequation*}
meaning that now the message (namely process $\phl P$) has finally been
delivered and it can be directly linked to $\phl R$ with a new (but smaller) \tsc{MCutQ}.


\noindent {\it Units ($\perp$ and $\one$).} These cases are a
simplified version of $\parr$ and $\tensor$ respectively:
\begin{smallequation*}
  \begin{array}{ll}
    \infer[\tsc{MCutQ}]{
      \phl{\res{x:\close x }(\wait xP)} \cll \Delta
    }
	  {
	    \infer[\bot]
		  {\phl {\wait xP} \cll \thl \Delta, \phl x:\thl\bot}
		  {\phl P \cll \thl\Delta}
		  &
		  \infer[\one]{
		    \close x \seq \pairQ{x}{\one^{\tilde x}}, \set[i]{\br[x]{\Star}\pairQ{x_i}{\cdot}}
		  }{ }
	  }
          & 
          \qquad\qquad\Longrightarrow\quad 
          \phl P \cll \thl\Delta
  \end{array}
\end{smallequation*}
\begin{smallequation*}
  \begin{array}{ll}
    \infer[\tsc{MCutQ}]{
      \phl{\res{x\tilde x:\wait xQ\br{\tilde y \lhd \tilde P}}(\close x \pp \tilde R)} 
      \cll 
      \set[j\le m]{\Delta_j}, \set[i\le n]{\Sigma_i}
    }{
	    \infer[\one]{\phl{\close x} \cll \phl x: \thl\one}{}
		  &
		  \quad\ldots\quad
		  &
		  \infer[\bot]{
		    \wait xQ \seq 
		    \dbr{\Psi_x}\pairQ{x}{\bot^{x_k}}, 
		    \Gamma
		  }{
		    \phl{Q} \seq 
		    \dbr{\Psi_x}\br[x_k]{\Star}\pairQ{x}{\cdot}, 
		    \Gamma
		  }
	  }
          \\[2ex]
          \qquad\Longrightarrow\qquad
	  \infer[\tsc{MCutQ}]{
		\phl{\res{x\tilde x:Q\br{\tilde y \lhd \tilde P}}(\tilde R)} \cll \set[j\le m]{\Delta_j}, \set[i\le n]{\Sigma_i}
	}{
	  \quad\ldots\quad
	  &
	  \phl{Q} \seq 
	  \dbr{\Psi_x}\br[x_k]{\Star}\pairQ{x}{\cdot}, 
	  \Gamma
	}
\end{array}
\end{smallequation*}

\noindent{\it Axiom.} Finally, the axiom (only defined on atoms, see~\cite{CLMSW16} for a discussion on eta-expansion) 
can completely erase a cut as follows:
\begin{smallalign*}
	&
  \infer[\tsc{MCutQ}]{ \phl{\res{xy:\fwd {x}{y}}(\fwd
      {x}{z} \pp \fwd {y}{w})} \cll \pairQ {z}{\dual a},
    \pairQ {w}{a} } { \infer[\tsc{Ax}]{ \fwd
      {x}{z1}\cll\pairQ{x}{a}, \pairQ {z}{\dual a} }{} &
    \infer[\tsc{Ax}]{ \fwd {y}{w} \cll\pairQ{y}{\dual a}, \pairQ
      {w}{a} }{} & \infer[\tsc{Ax}]{ \fwd {x}{y}
      \seq\pairQ{x}{\dual a}, \pairQ {y}{a} }{} }
\\	&
\qquad\Longrightarrow\qquad
  \infer[\tsc{Ax}]{
    \fwd {z}{w}\cll\pairQ {z}{\dual a}, \pairQ {w}{a}
  }{}
\end{smallalign*}

These reductions allow us to prove the key lemma of this section.
\begin{lemma}[Admissibility of $\tsc{MCutQ}$]
  If
  $\set[j\le m]{\phl {P_j} \cll \thl\Delta_j, \pairQ{y_j}{A_j}}$
  and
  
  \noindent
  $\set[i\le n]{\phl {R_i} \cll \thl\Sigma_i, \pairQ{x_i}{B_i}}$
  and 
  $\phl Q \seq \set[i\le n]{\dbr{\Psi_i}\pairQ{x_i}\dual{B_i}}, \set[n < i \le p]{ \dbr{\Psi_i}\pairQ{x_i}{\cdot}}$
  then there exists a process $\phl S$ such that   
  $\phl{\res{\tilde x: Q \br{\tilde y \lhd \tilde P}} \tilde R} \Rightarrow^* \phl S$
  and
  $\phl S \cll \set[j\le m]{\Delta_j}, \set[i\le n]{\Sigma_i}$.
\end{lemma}
\begin{proof}[Sketch]
	By lexicographic induction on ($i$) the sum of sizes of the $B_i$'s and ($ii$) the sum of  sizes of the $\phl{R_i}$'s.
	The base cases and key cases have been detailed above.
	The commutative cases are straightforward and only need to consider the possible last rule applied to a premiss of the form $\phl {R_i} \cll \thl\Sigma_i, \pairQ{x_i}{B_i}$. \qed
\end{proof}

We can finally conclude with the following theorem as a special case.

\begin{theorem}[Admissibility of $\tsc{MCutF}$]
	If
	$\set[i\le n]{\phl {R_i} \cll \thl\Sigma_i, \pairQ{x_i}{A_i}}$
	and
	$\phl Q \seq \thl\set[i\le n]{\pairQ{x_i}\dual{A_i}}$
	then there exists a process $\phl S$ such that   
	$\phl{\res{\tilde x:Q}(R_1 \pp \ldots \pp R_n)}\Rightarrow^* \phl S$
	and
	$\phl S \cll \set[j\le m]{\Delta_j}, \set[i\le n]{\Sigma_i}$.
\end{theorem}

%
%

\section{Asynchronous Forwarders as Multiparty
  Compatibility}\label{sec:mcompatibility}
In the previous section, we have shown that our forwarders can be used
to govern the composition of processes. In this section, we show that
forwarders precisely capture the intuitive notion of compatibility,
i.e., {\em given a set of processes, is there a communication pattern
  they can follow so that they progress without reaching an error?}
The concept of compatibility has been studied in the context of
multiparty session types~\cite{LY19,GPPSY21}. Below, we adapt
multiparty compatibility to our logical setting, focusing on CP
processes.  We start by introducing the concept of semantics for type
contexts:
\begin{definition}[Type-Context Semantics]
  Let $\Gamma$ be a forwarder type context. Then, we define
  $\lts{\alpha}$ as the minimum relation satisfying the following
  rules:
  \begin{smallequation*}
    \begin{array}{lc@{\quad}l}
      \Gamma, \dbr{\Psi}\pairQ x{A \parr^{\phl u} B}
      &\lts{\phl x\tensor\phl u}&
      \Gamma, \dbr{\Psi}\br[u]{\pairQ yA}\pairQ xB
      \\
      \Gamma,
      \br[x]{\pairQ{z}{\dual A}}\dbr{\Psi_u}\pairQ{u}{C},
      \dbr{\Psi_x}\pairQ x{A \tensor^{\phl u} B}
      &\lts{\phl x\parr \phl u}&
      \Gamma, \dbr{\Psi_u}\pairQ{u}{C}, \dbr{\Psi_x}\pairQ xB
      \\
      \Gamma, \dbr{\Psi}\pairQ{x}{\bot^\phl{u}}
      &\lts{\phl x\one \phl u}&
      \Gamma, \dbr{\Psi}\br[u]{\Star}\pairQ{x}{\cdot}
      \\
      \set[1\le i \le n]{\br[x]{\Star}\pairQ{u_i}{\cdot}}, \pairQ{x}{\one^\phl{\tilde u}}
      &\lts{\phl x\perp \phl {\tilde u}}&
      \checkmark
      \\
      \pairQ{x}{\dual a}, \pairQ ya
      &\lts{\phl x\fwd \phl y}&
      \checkmark
    \end{array}
  \end{smallequation*}
\end{definition}
The rules above capture an asynchronous semantics for typing contexts
(for the multiplicative fragment of CP). We observe that our
definition, despite looking different, is equivalent to that given by
Ghilezal et al.~\cite{GPPSY21}. In fact, our context uses a dualised
version: in order to obtain their setting, we should just consider a
double context containing the dual of the endpoint types and the dual
of the queues, respectively. Also note that since we have no infinite
computations, we do not need to consider fairness.
Using the relation on contexts above, we can define when a set of
endpoints successfully progresses without reaching an error.  This can
be formalised by the concept of live path. In the sequel, let $\alpha$
range over all the possible labels of the relation above. Moreover,
let $\alpha_1,\ldots,\alpha_n$ be a {\em path} for a context $\Gamma$
whenever there exist $\Gamma_1, \ldots,\Gamma_n$ such that
$\Gamma\lts{\alpha_1}\Gamma_1\ldots\lts{\alpha_n}\Gamma_n$.
\begin{definition}[Live Path]
  Let $\Gamma$ be a context and let $\alpha_1,\ldots,\alpha_n$ be a
  path for $\Gamma$. We say that the path $\alpha_1,\ldots,\alpha_n$ is
  {\em live} if $\Gamma\lts{\tilde\alpha}\checkmark$ and, if for some $i< n$, 
  \begin{enumerate}
  \item\label{one} 
    $\Gamma_i=\Gamma_i', \br[x]{\pairQ{z}{\dual
        A}}\dbr{\Psi_u}\pairQ{u}{C}$, then there exists $k > i$ 
    such that  $\alpha_k = \phl x\parr \phl u$;
   
     \item\label{two} 
    $\Gamma_i=\Gamma_i', \br[x]{\Star}\pairQ{u}{\cdot}$, 
    then   
    $\alpha_n=\phl x\perp \phl {\tilde u}$ with
    $\phl u\in\phl{\tilde u}$;
    
    \item\label{three} 
    $\Gamma_i = \Gamma_i', \dbr{\Psi_x}\pairQ x{A\tensor^{\phl u} B}$,
    then there exists $k> i$ such that
    $\alpha_k = \phl x\parr \phl u$;
    
  \item\label{four} 
  $\Gamma_i = \Gamma_i', \dbr{\Psi_x}\pairQ x{\one^\phl{\tilde u}}$,
  then 
  $\alpha_n=\phl x\perp \phl {\tilde u}$.

  \end{enumerate}
\end{definition}
Conditions~\eqref{one} and~\eqref{two} 
state that every message that has
been enqueued in the forwarder is eventually forwarded. 
On the other
hand, conditions~\eqref{three} and ~\eqref{four} state that every
forwarding instruction is eventually executed. 
These conditions are
basically what our forwarders do, as stated by the following:
\begin{theorem}
$\Gamma$ has a live path iff
  there exists a forwarder $\phl F$ such that $\phl F\seq\Gamma$.
\end{theorem}
\begin{proof} We need to prove each direction separately. For the
  if direction, we proceed by induction on the proof of
  $\phl F$ (logically, on its $\eta$-long normal form). For the only-if
  direction, we proceed by induction on the length of the path. \hfill $\Box$
\end{proof}

%
%

\section{Related Work}\label{sec:related}
Our work takes~\cite{CLMSW16} as a starting point.
Guided by CLL, we set out to explore if coherence can be broken down
into more elementary logical rules which led us to introduce
forwarders. As a result, forwarders provide a more general notion of
compatibility.
An earlier unpublished version of this work~\cite{CMS21}, proposes
synchronous forwarders, i.e., the restriction of forwarders with only
buffers of size one.
In that case, we show that we can always construct a coherence proof
from a synchronous forwarder.  However, synchronous forwarders fail to
capture all the possible interleaving of an arbiter (encoding of
coherence to processes).

Caires and Perez~\cite{CP16} also study multiparty session types in
the context of intuitionistic linear logic by translating global types
to processes, called \emph{mediums}.
Their work does not start from a logical account of global types
(their global types are just syntactic terms). But, as previous
work~\cite{CLMSW16}, they do generate arbiters as linear logic proofs,
which are special instances of forwarders. In this work, we generalise
this approach to characterise exactly which processes can justify the
compatibility of several processes.
%

%

Sangiorgi~\cite{S96}, probably the first to treat forwarders for the
$\pi$-calculus, uses binary forwarders, i.e., processes that only
forward between two channels, which are equivalent to our $\fwd xy$. We
attribute our result to the line of work that originated in 2010 by
Caires and Pfenning~\cite{CP10},
where forwarders \emph{\`a la Sangiorgi} were introduced as processes
to be typed by the axiom rule in linear logic. Van den Heuvel and
Perez~\cite{Heuvel2020SessionTS} have recently developed a version of
linear logic that encompasses both classical and intuitionistic logic,
presenting a unified view on binary forwarders in both logics.

Gardner et al.~\cite{GLW07} study the expressivity of the linear
forwarder calculus, by encoding the asynchronous $\pi$-calculus (since
it can encode distributed choice).  The linear forwarder calculus is a
variant of the (asynchronous) $\pi$-calculus that has binary
forwarders and a restriction on the input $x(y).P$ such that $y$
cannot be used for communicating (but only for forwarding). Such a
restriction is similar to the intuition behind our forwarders,
with the key difference that their methodology would not apply to some of our
session-based primitives. 

Barbanera and Dezani~\cite{BD19} study multiparty session types as
\emph{gateways} (which are basically forwarders) that work as a medium among
many interacting parties, forwarding communications between two
multiparty sessions. Such mechanism reminds us of our forwarder
composition: indeed, in their related work discussion they do mention
that their gateways could be modelled by 
a ``connection-cut''.

Recent work~\cite{JGP16,GJP18} proposes an extension of linear logic
that models {\em identity providers}, a sort of monitoring mechanisms
that are basically forwarders between two channels in the sense of
Sangiorgi, but asynchronous, i.e., they allow unbounded buffering of
messages before forwarding. Our forwarders can be seen as a
generalisation to multiparty monitors. Multiparty monitors are also
addressed by Hamers and Jongmans~\cite{HJ20}, but not in a linear
logic context.

Our forwarder mechanism may be confused with that of
locality~\cite{MS04}, which is discussed from a logical point of view
by Caires et al.~\cite{CPT16}. Locality only requires that received
channels cannot be used for inputs (which then must occur at the
location where the channel was created). 
In our case instead, we do not allow received channels to be used at
all until a new forwarder is created.

\section{Conclusions}\label{sec:conclusions}
In this paper, we have presented a logical characterisation of forwarders and 
their semantics based on cut elimination.
Additionally, we have shown that 
forwarders can replace the notion of coherence when composing multiple
processes; indeed, forwarders can implement all and only compatible
multiparty communications within linear logically typable sessions.
 We discuss some aspects of forwarders developed in this paper and
identify possible future extensions. 



\mypar{Process Language.} Our process language is based on~\cite{W14} with some omissions. For the sake of presentation, we
have left out the additive and exponential fragment (which can be
found in the appendix). Moreover, we have not included polymorphic
communications. We show in the appendix that the logic presented in this paper extends
directly to additives and exponentials, and we conjecture that our forwarder logic extends to polymorphic types $\exists X.A$ and $\forall X.A$ as well. A further extension to support recursion
is left to future work.

\mypar{Classical vs.~Intuitionistic Linear Logic.} In this paper, we
have chosen to base our theory on CLL for two main reasons.  Coherence
is indeed defined by Carbone et al.~\cite{CLMSW16} in terms of CLL and
therefore our results can immediately be related to theirs without
further investigations. 
%
An earlier version of the forwarder logic was based on intuitionistic
linear logic, but moving to CLL required fewer rules and greatly
improved the presentation.  Nevertheless, our results should be easily
reproducible in intuitionistic linear logic.


\mypar{Variants of Coherence.} Our results show that forwarders are a
generalisation of coherence proofs.  Indeed, coherence would
correspond to the notion of \emph{synchronous
  forwarders}~\cite{CMS21}, the restriction of forwarders with only
buffers of size one.  As a follow-up, we would like to investigate,
whether other syntactic restrictions of forwarders also induce
interesting generalised notions of coherence, and, as a consequence,
generalisations of global types.

%
%

%
%
%
\bibliographystyle{alpha}
\bibliography{biblio}

\newpage
\appendix

\section{Full CP and Classical Linear Logic} \label{app:cll}
\mypar{Types.}
\begin{smallequation}\label{eq:types}
	\thl{A} ::=  \quad
	\thl{a}
	\: \mid\: \thl{a^\perp}
	\: \mid\: \thl{1}
	\: \mid\: \thl{\bot}
	\: \mid\: \thl{(A\tensor A)}
	\: \mid\: \thl{(A\parr A)}
	\: \mid\: \thl{(A\oplus A)}
	\: \mid\: \thl{(A\with A)}
	\: \mid\: \thl{!A}
	\: \mid\: \thl{?A} 
\end{smallequation}
\mypar{Duality.}
\begin{smallequation*}
	\dual{(\dual{a})} = a
	\quad \dual\one=\perp
	\quad \dual{(A\tensor B)} = \dual A\parr\dual B
	\quad \dual {(A\oplus B)}=\dual A\with\dual B
	\quad \dual{(!A)} = ?\dual A
\end{smallequation*}

\mypar{Processes.} 
\begin{smallequation*}
  \begin{array}{l@{\quad}l@{\quad}l@{\qquad}l@{\quad}l@{\quad}l@{\qquad}l}
    \phl P,\phl Q ::= & \fwd xy          & \text{(link)}
    &
      \phl{\res {xy} (P\!\!\pp\!\! Q)} & \text{(parallel)}
    \\
                      & \wait x P & \text{(wait)}
    &
      \close x & \text{(close)}
    \\
                      & \recv xyP        & \text{(input)}
                
    &
      \send xyPQ       & \text{(output)}
    \\
            & \Case xPQ        & \text{(choice)}
    & \inl xP          & \text{(left select)} \\
& & & \inr xP & \text{(right select)}
    \\
            & \client xyP & \text{(client request)}
    &
      \srv xyP & \text{(server accept)}
  
  \end{array}
\end{smallequation*}

\mypar{CP-typing.}
\begin{smallequation*}
    \begin{array}{c}
      \infer[\textsc{Ax}]
      {\phl{\fwd{x}{y}} \cll \phl x:\thl {a^\bot}, \phl y:\thl{a}}
      { }
      \qquad
      \infer[\bot]
      {\phl {\wait xP} \cll \thl \Delta, \phl x:\thl\bot}
      {\phl P \cll \thl\Delta}
      \qquad
      \infer[\one]
      {\phl{\close x} \cll \phl x: \thl\one}
      { }
      \\[1ex]
      \infer[\tensor]
      {
      \phl{\send xyPQ} 
      \cll 
      \thl{\Delta_1}, \thl{\Delta_2}, \phl x:\thl{A_1 \tensor A_2}
      }
      {
      \phl P 
      \cll 
      \thl{\Delta_1}, \phl y:\thl A_1
      & \phl Q 
        \cll 
        \thl \Delta_2, \phl x: \thl A_2
        }
        \qquad\qquad\qquad
        \infer[\parr]
        {\phl{\recv xyP} 
        \cll \thl{\Delta}, \phl x: \thl {A_1 \parr A_2}
        }       
        {\phl P \cll \thl{\Delta}, \phl y:\thl A_1, \phl x:\thl A_2}
      \\[1ex]
      \infer[\oplus_1]
      {\phl{\inl xP} \cll \thl\Delta, \phl x:\thl{A_1 \oplus A_2}}
      {\phl P \cll \thl\Delta, \phl x:\thl A_1}
      \qquad
      \infer[\oplus_2]
      {\phl{\inr xP} \cll \thl\Delta, \phl x:\thl{A_1 \oplus A_2}}
      {\phl P \cll \thl\Delta, \phl x:\thl A_2}
      \qquad
      \infer[\with]
      {\phl{\Case xPQ} \cll \thl \Delta, \phl x:\thl{A_1 \with A_2}}
      {\phl P \cll \thl\Delta, \phl x:\thl A_1 
      & 
        \phl Q \cll \thl\Delta, \phl x:\thl A_2}
      \\[1ex]
      \infer[?]
      {\phl{\client xyP} \cll \thl \Delta, \phl x:\thl {\query A}}
      {\phl P \cll \thl \Delta, \phl y:\thl A}
      \qquad\qquad
      \infer[!]
      {\phl{\srv xyP} \cll \,\thl{\query \Delta}, \phl x:\thl{\bang A}}
      {\phl P \cll \,\thl{\query \Delta}, \phl y:\thl A}
      \\[1ex]
      \infer[\textsc{Weaken}]
      {\phl P \cll \thl \Delta, \phl x:\thl{\query A}}
      {\phl P \cll \thl \Delta}
                                                 \ \quad
                                                 \infer[\textsc{Contract}]
                                                 {\phl{P\{x/y, x/z\}} \cll \thl\Delta, \phl x:\thl{\query A}}
                                                 {\phl P \cll \thl\Delta, \phl y:\thl{\query A}, \phl
                                                 z:\thl{\query A}}
    \end{array}
  \end{smallequation*}

\section{Rules -- including gathering and broadcasting}
We can complexify the types in order to allow broadcasting and gathering.
\[
\begin{array}{rcllllllll}
	\thl{A,B} & \coloncolonequals & 
	\thl{a}
	\quad\mid\quad \thl{a^\perp} 
	\\ &
	\quad\mid\quad & \thl{\one^{[u_1, \ldots, u_n]}}&
	\quad\mid\quad &\thl{\bot^u} &
	\\ &
	\quad\mid\quad &\thl{(A\tensor^{[u_1,\ldots,u_n]} B)} &
	\quad\mid\quad & \thl{(A\parr^u B)} 
	\\ &
	\quad\mid\quad & \thl{(A\oplus^u B)} &
	\quad\mid\quad & \thl{(A\with^{[u_1, \ldots, u_n]} B)} 
	\\ &
	\quad\mid\quad & \thl{?^u A} &
	\quad\mid\quad & \thl{!^{[u_1, \ldots, u_n]} A} 
\end{array}
\]

Contexts need to be changed as well.
\[
\begin{array}{rcllllllll}
	\dbr \Psi & \coloncolonequals & \cdot 
	& \quad\mid\quad \br[u]{\Star}
	& \quad\mid\quad \br[u]{\pairQ yB}
	& \quad\mid\quad \dbr \Psi \dbr \Psi 
	\\ &&
	& \quad\mid\quad \br[u]{\Left}
	& \quad\mid\quad \br[u]{\Right}
	& \quad\mid\quad \br[u]{\Query}
\end{array}
\]
%

$\br[u]{\Left}$ (or $\br[u]{\Right}$) and
$\br[u]{\Query}$ indicate that 
a branching request and  server invocation, respectively, has been
received and must be forwarded

When it needs to be forwarded to a list of endpoints,
we may use the abbreviation $\br[\tilde u]{\Symbol}$ for denoting the
$\br[u_1]{\Symbol}\ldots\br[u_n]{\Symbol}$ when
$\phl{\tilde u} = \phl{u_1}, \ldots, \phl{u_n}$.
In this case, we also assume the implicit rewriting
$\br[\emptyset]{\Symbol}\dbr{\Psi} \equiv \dbr{\Psi}$.

The rules then need to be adapted as follows.
\begin{displaymath}
	\begin{array}{c}
		\infer[\tsc{Ax}]{
			\fwd xy\seq\pairQ{x}{\dual a}, \pairQ ya
		}{}
		\qquad
		\infer[\bot]{
			\wait xP \seq \Gamma, \dbr{\Psi}\pairQ{x}{\bot^\phl{u}}
		}{
			\phl{P} \seq \Gamma, \dbr{\Psi}\br[u]{\Star}\pairQ{x}{\cdot}
		}
		\qquad	
		\infer[\one]{
			\close x \seq \set[i]{\br[x]{\Star}\pairQ{u_i}{\cdot}}, \pairQ{x}{\one^{\tilde u}}
		}{ }
		\\\\
		\infer[\parr]{
			\phl{\recv xyP} 
			\seq 
			\Gamma, \dbr{\Psi}\pairQ x{A \parr^\phl{u} B} 
		}{
			\phl P 
			\seq 
			\Gamma, \dbr{\Psi}\br[u]{\pairQ yA}\pairQ xB
		}
		\qquad
		\infer[\tensor\ (\Delta_i \not= \emptyset)]{
			\phl{\send xy{P}Q} 
			\seq 
			\Gamma,
			\big\{\br[\phl x]{\Delta_i}\dbr{\Psi_i}\pairQ{u_i}{C_i}\big\}_i,
			\dbr{\Psi}\pairQ x{A \tensor^{\phl{\tilde u}} B}
		}{
			\phl {P}
			\seq
			\big\{\Delta_i\big\}_i,\pairQ yA
			& 
			\phl Q 
			\seq
			\Gamma, \big\{\dbr{\Psi_i}\pairQ{u_i}{C_i}\big\}_i,
			\dbr{\Psi}\pairQ xB
		}
		%
		%
		\\\\
		\infer[\with\ (\tilde u \not= \emptyset)]{
			\Case{x}{P}{Q}
			\seq \Gamma, \dbr{\Psi}\pairQ x{A \with^\phl{\tilde u} B}
		}{
			\phl P\seq \Gamma, \dbr{\Psi}\br[\tilde u]{\Left}\pairQ xA
			&
			\phl Q\seq \Gamma, \dbr{\Psi}\br[\tilde u]{\Right}\pairQ xB
		}
		\\\\
		\infer[\oplus_l]{
			\inl xP\seq \Gamma, \br[x]{\Left}\dbr{\Psi_z} \phl z: \thl C,
			\dbr{\Psi_x}\pairQ x{A \oplus^\phl{z} B}
		}{
			\phl P\seq \Gamma, \dbr{\Psi_z} \phl z: \thl C, \dbr{\Psi_x}\pairQ xA
		}
		%
		\qquad
		\infer[\oplus_r]{
			\inr xP\seq \Gamma, \br[x]{\Right}\dbr{\Psi_z} \phl z: \thl C,
			\dbr{\Psi_x}\pairQ x{A \oplus^\phl{z} B}
		}{
			\phl P\seq \Gamma, \dbr{\Psi_z} \phl z: \thl C, \dbr{\Psi_x}\pairQ xB
		}
		%
		\\\\
		\infer[!\ (\tilde u \not= \emptyset)]{
			\srv xyP \seq \set[i]{\pairQ{u_i}{\query B_i}}, \pairQ x{\bang^\phl{\tilde u} A} 
		}{
			\phl{P} \seq \set[i]{\pairQ{u_i}{\query B_i}}, \br[\tilde u]{\Query}\pairQ y A
		}
		\qquad
		\infer[?]{
			\client xyP \seq \Gamma, \br[x]{\Query}\dbr{\Psi_z}\pairQ zC, \dbr{\Psi_x}\pairQ x{\query^\phl{z} A}
		}{
			\phl{P}\seq \Gamma, \dbr{\Psi_z}\pairQ zC, \dbr{\Psi_x}\pairQ yA
		}
		%
	\end{array}
\end{displaymath} 

Proposition~\ref{prop:embed} is naturally extended to additives and exponentials.

\begin{proposition}\label{prop:embed2}
	Any forwarder is a CP-typable process, that is, if $\phl P \seq \Gamma$, then  $\phl P \cll \enc{\Gamma}$.
	The embedding $\enc{\cdot}$ being extended as: $\enc{\br[u]{\Symbol}\dbr{\Psi}} = \enc{\dbr{\Psi}}$.
\end{proposition}

\section{Cut admissibility}
\subsection{General distribution and substitution}

\mypar{Distribution}

%
%

Multiplicatives with gathering
\[
		\left\{
		\begin{array}{l}
			\Gamma_1, \dbr{\Psi_{2z}} \bullet \dbr{\Psi_z} \pairQ
			z{B\{\tensor^\phl{\tilde vx}\}}
			\rhd
			\br[\phl z]{\{\Delta_i\}_i}\dbr{\Psi_1}\pairQ xA
			\\
			\Gamma_2, \big\{
			\dbr{\Psi_{1c_i}} \bullet \dbr{\Psi_{c_i}} \pairQ
			{c_i}{C_i}  \big\}_i
			\rhd
			\dbr{\Psi_2}\pairQ y{\dual A}
		\end{array}
		\right\}
		\xrightarrow{}_{\mathsf{distr}} 
\]
\[
		\left\{
		\begin{array}{l}
			\Gamma_1, \dbr{\Psi_{2z}} \bullet \dbr{\Psi_z} \pairQ
			z{B\{\tensor^\phl{\tilde v\tilde c}\}}
			\rhd
			\dbr{\Psi_1}\pairQ xA
			\\
			\Gamma_2, \big\{
			\dbr{\Psi_{1c_i}}  \br[\phl z]{\Delta_i}\bullet \dbr{\Psi_{c_i}} \pairQ
			{c_i}{C_i}  \big\}_i
			\rhd
			\dbr{\Psi_2}\pairQ y{\dual A}
		\end{array}
		\right\}
\]

Additives

$$
\left\{
\begin{array}{rcl}
	\Gamma_1, \dbr{\Psi_{2d}}\bullet\dbr{\Psi_d}\pairQ{d}{D\ctx{\oplus^{\bhl{\bm x}}}} & \rhd & \br[d]{\Left}\dbr{\Psi_1}\phl x: \thl A \\
	\Gamma_2, \dbr{\Psi_{1c}}\bullet\dbr{\Psi_c}\pairQ{c}{D} & \rhd & \dbr{\Psi_2}\pairQ{y}{\dual A}
\end{array}
\right\}
\xrightarrow{}_{\mathsf{distr}} 
$$

$$
\left\{
\begin{array}{rcl}
	\Gamma_1, \dbr{\Psi_{2d}}\bullet\dbr{\Psi_d}\pairQ{d}{D}\{\oplus^\bhl{\bm c}\} & \rhd &\dbr{\Psi_1}\pairQ{x}{A} \\
	\Gamma_2, \dbr{\Psi_{1c}}\br[d]{\Left}\bullet\dbr{\Psi_c}\pairQ{c}{C} & \rhd & \dbr{\Psi_2}\pairQ{y}{\dual A}
\end{array}
\right\}
$$

Exponentials 

$$
\left\{
\begin{array}{rcl}
	\Gamma_1, \dbr{\Psi_{2d}}\bullet\dbr{\Psi_d}\pairQ{d}{D\ctx{\query^{\bhl{\bm x}}}} & \rhd & \br[d]{\Query}\dbr{\Psi_1}\phl x: \thl A \\
	\Gamma_2, \dbr{\Psi_{1c}}\bullet\dbr{\Psi_c}\pairQ{c}{D} & \rhd & \dbr{\Psi_2}\pairQ{y}{\dual A}
\end{array}
\right\}
\xrightarrow{}_{\mathsf{distr}} 
$$

$$
\left\{
\begin{array}{rcl}
	\Gamma_1, \dbr{\Psi_{2d}}\bullet\dbr{\Psi_d}\pairQ{d}{D}\{\query^\bhl{\bm c}\} & \rhd &\dbr{\Psi_1}\pairQ{x}{A} \\
	\Gamma_2, \dbr{\Psi_{1c}}\br[d]{\Query}\bullet\dbr{\Psi_c}\pairQ{c}{C} & \rhd & \dbr{\Psi_2}\pairQ{y}{\dual A}
\end{array}
\right\}
$$

\mypar{Substitution}

%

Multiplicatives with gathering
\[
		\left\{
		\begin{array}{lcl}
			\Gamma_1, \dbr{\Psi_z}\pairQ
			z{E\{\tensor^{\phl{\tilde vx}}\}}
			&\rhd&
			\pairQ x{A\parr^{\phl z} B}
			\\
			\Gamma_2, \big\{\br[y]{\Delta_i}\dbr{\Psi_i}] \pairQ{c_i}{C_i}  \big\}_i,
			\big\{\dbr{\Psi_j}\pairQ{d_j}{D_j\ctx{\parr^{\phl y}}}\big\}_j
			&\rhd&
			\pairQ y{\dual A\tensor^\phl{\tilde c\tilde d}\dual B}
		\end{array}
		\right\}
		\xrightarrow{}_{\mathsf{subst}} 
\]
\[	
\left\{
\begin{array}{lcl}
			\Gamma_1, \dbr{\Psi_z}\pairQ z{E\ctx\tensor^{\phl{\tilde v\tilde c\tilde d}}}
			&\rhd&
			\pairQ x{B}
			\\
			\Gamma_2, \big\{\br[z]{\Delta_i}\dbr{\Psi_i} \pairQ{c_i}{C_i}  \big\}_i,
			\big\{\dbr{\Psi_j}\pairQ{d_j}{D_j\{\parr^{\phl z}\}}\big\}_j
			&\rhd&
			\pairQ y{\dual B}
		\end{array}
		\right\}
\]

Additives
\[
\left\{
\begin{array}{lcl}
	\Gamma_1, \big\{\dbr{\Psi_{u_i}}\pairQ{u_i}{C_i\ctx{\oplus^\phl{x}}}\big\}_i
	&\rhd &
	\pairQ x{A \with^\phl{\tilde u} B}
	\\
	\Gamma_2, \br[y]{\Left}\dbr{\Psi_v} \phl v: \thl D
	&\rhd &
	\pairQ y{\dual A \oplus^\phl{v} \dual B}
\end{array}
\right\}
\longrightarrow_{\mathsf{subst}}
\left\{
\begin{array}{lcl}
	\Gamma_1, \big\{\dbr{\Psi_{u_i}}\pairQ{u_i}{C_i\ctx{\oplus^\phl{v}}}\big\}_i
	&\rhd &
	\pairQ x{B}
	\\
	\Gamma_2, \br[\tilde u]{\Left}\dbr{\Psi_v} \phl v: \thl D
	&\rhd &
	\pairQ y{\dual B}
\end{array}
\right\}
\]

\[
\left\{
\begin{array}{lcl}
	\Gamma_1, \big\{\dbr{\Psi_{u_i}}\pairQ{u_i}{C_i\ctx{\oplus^\phl{x}}}\big\}_i
	&\rhd &
	\pairQ x{A \with^\phl{\tilde u} B}
	\\
	\Gamma_2, \dbr{\Psi_v} \phl v: \thl D\ctx{\with^\phl{y\tilde z}}
	&\rhd &
	\pairQ y{\dual A \oplus^\phl{v} \dual B}
\end{array}
\right\}
\longrightarrow_{\mathsf{subst}}
\left\{
\begin{array}{lcl}
	\Gamma_1, \big\{\dbr{\Psi_{u_i}}\pairQ{u_i}{C_i\ctx{\oplus^\phl{v}}}\big\}_i
	&\rhd &
	\pairQ x{B}
	\\
	\Gamma_2, \dbr{\Psi_v} \phl v: \thl D\ctx{\with^\phl{\tilde u\tilde z}}
	&\rhd &
	\pairQ y{\dual B}
\end{array}
\right\}
\]

Exponentials

\[
\left\{
\begin{array}{lcl}
	\Gamma_1, \set[i]{\dbr{\Psi_{u_i}}\pairQ{u_i}{B_i\ctx{\query^\phl{x}}}}
	&\rhd &
	\pairQ x{\bang^\phl{\tilde u} A} 
	\\
	\Gamma_2, \br[y]{\Query}\dbr{\Psi_z}\pairQ z{C}
	&\rhd &
	\pairQ y{\query^\phl{z} \dual A}
\end{array}
\right\}
\longrightarrow_{\mathsf{subst}}
\left\{
\begin{array}{lcl}
	\Gamma_1, \set[i]{\dbr{\Psi_{u_i}}\pairQ{u_i}{B_i\ctx{\query^\phl{z} }}}
	&\rhd &
	\pairQ x{A} 
	\\
	\Gamma_2, \br[\tilde u]{\Query}\dbr{\Psi_z}\pairQ z{C}
	&\rhd &
	\pairQ y{\dual A}
\end{array}
\right\}
\]

\[
\left\{
\begin{array}{lcl}
	\Gamma_1, \set[i]{\dbr{\Psi_{u_i}}\pairQ{u_i}{B_i\ctx{\query^\phl{x}}}}
	&\rhd &
	\pairQ x{\bang^\phl{\tilde u} A} 
	\\
	\Gamma_2, \dbr{\Psi_z}\pairQ z{C\ctx{\bang^\phl{y}}}
	&\rhd &
	\pairQ y{\query^\phl{z} \dual A}
\end{array}
\right\}
\longrightarrow_{\mathsf{subst}}
\left\{
\begin{array}{lcl}
	\Gamma_1, \set[i]{\dbr{\Psi_{u_i}}\pairQ{u_i}{B_i\ctx{\query^\phl{z} }}}
	&\rhd &
	\pairQ x{A} 
	\\
	\Gamma_2, \dbr{\Psi_z}\pairQ z{C\ctx{\bang^\phl{\tilde u}}}
	&\rhd &
	\pairQ y{\dual A}
\end{array}
\right\}
\]

\subsection{Cut-in-box lemmas}

%

\begin{lemma} 
	\label{lem:cuttwo-adm}
  Let $\phl P\seq\Delta, \pairQ xA$ and
  $\phl Q\seq\Gamma, \dbr{\Psi_1}\br[\phl z]{\pairQ w{\dual A}}\dbr{\Psi_2}\pairQ y{B}$ be derivable.
  Then, there exists $\phl R$ such that
  $\phl R\seq^\cut \Gamma, \dbr{\Psi_1}\br[\phl z]{\Delta}\dbr{\Psi_2}\pairQ y{B}$ is
  derivable with $\cut$ and $\mathtt{rank}(\phl{R}) = \mathtt{size}(A)$.
  We denote this process as $\phl R = \phl{\res{x\rhd w}(P\pp Q)}$
\end{lemma}
\begin{proof}
By induction on the structure of $\phl Q$.
Because of the shape of its typing environment, $\phl Q$ cannot be of the shape $\fwd ab$ nor of the shape $\close a$.

Case $\phl Q = \send zvST$ and $\phl w\in \mathtt{fn}(\phl S)$.
Assume we have the following
\[
\infer[\tensor]{
	\phl{\send zvST} 
	\seq 
	\Gamma', \br[z]{\pairQ w{\dual A}}\dbr{\Psi_y}\pairQ yB, 
	\big\{\br[\phl z]{\Delta_i}\dbr{\Psi_{u_i}}\pairQ{u_i}{C_i}\big\}_i, \dbr{\Psi_z}\pairQ z{D \tensor^\phl{y \tilde u} E}
}{
	\phl{S} \seq 
	\pairQ w{\dual A}, \big\{\Delta_i\big\}_i, \pairQ vD
	&
	\phl{T} \seq
	\Gamma', \dbr{\Psi_y}\pairQ yB, \big\{\dbr{\Psi_{u_i}}\pairQ{u_i}{C_i}\big\}_i, \dbr{\Psi_z}\pairQ zE
}
\]
%
We take $\phl R = \send zv{\res{xw}(P\pp S)}T$ and indeed $\mathtt{rank}(\phl R) = \mathtt{size}(A)$ as follows:
\[
\infer[\tensor]{
	\send zv{\res{xw}(P\pp S)}T
	\seq
	\Gamma', \br[z]{\Delta}\dbr{\Psi_y}\pairQ yB, 
	\big\{\br[\phl z]{\Delta_i}\dbr{\Psi_{u_i}}\pairQ{u_i}{C_i}\big\}_i, \dbr{\Psi_z}\pairQ z{D \tensor^\phl{y \tilde u} E}
}{	
	\infer[\cut]{
		\phl{\res{xw}(P\pp S)} \seq \Delta, \big\{\Delta_i\big\}_i, \pairQ vD
	}{
		\phl{P} \seq \Delta, \pairQ xA
		&
		\phl{S} \seq \pairQ w{\dual A}, \big\{\Delta_i\big\}_i,\pairQ vD
	}
	&
	\phl{T} \seq
	\Gamma', \dbr{\Psi_y}\pairQ yB, \big\{\dbr{\Psi_{u_i}}\pairQ{u_i}{C_i}\big\}_i, \dbr{\Psi_z}\pairQ zE
}
\]

This works because the labels in the left premiss are irrelevant to the conclusion, since the $\Delta$'s and the $D$ in the conclusion are not annotated (inside boxes and antecedent of $\tensor$ resp.).

Case $\phl Q = \send zvST$ and $\phl w\in \mathtt{fn}(\phl T)$.
Assume we are in the following context 
\[
\infer[\tensor]{
	\send zvST
	\seq
	\Gamma', \br[z]{\Delta'}\dbr{\Psi_1}\br[z]{\pairQ w{\dual A}}\dbr{\Psi_2}\pairQ yB, 
	\big\{\br[z]{\Delta_i}\dbr{\Psi_{u_i}}\pairQ{u_i}{C_i}\big\}_i,  \dbr{\Psi_z}\pairQ z{C \tensor^\phl{y \tilde u} D}
}{
	\phl{S} \seq \Delta', \big\{\Delta_i\big\}_i,\pairQ vC
	&
	\phl T
	\seq
	\Gamma', \dbr{\Psi_1}\br[z]{\pairQ w{\dual A}}\dbr{\Psi_2}\pairQ yB, \big\{\dbr{\Psi_{u_i}}\pairQ{u_i}{C_i}\big\}_i,  \dbr{\Psi_z}\pairQ z{D}
}
\]
By applying the induction hypothesis to $\phl P$ and $\phl T$ (which indeed is a subprocess of $\phl Q$), we obtain $\phl{R_0}$ such that
$$\phl{R_0}\seq^\cut 
\Gamma', \dbr{\Psi_1}\br[z]{\Delta}\dbr{\Psi_2}\pairQ yB, \big\{\dbr{\Psi_{u_i}}\pairQ{u_i}{C_i}\big\}_i, \dbr{\Psi_z}\pairQ z{D}$$
and  $\mathtt{rank}(\phl{R_0}) = \mathtt{size}(A)$.
So we can take $\phl R = \send zvS{R_0}$
and indeed, 
$$\phl R \seq^\cut \Gamma', \br[z]{\Delta'}\dbr{\Psi_1}\br[z]{\Delta}\dbr{\Psi_2}\pairQ yB, 
\big\{\br[z]{\Delta_i}\dbr{\Psi_{u_i}}\pairQ{u_i}{C_i}\big\}_i,  \dbr{\Psi_z}\pairQ z{C \tensor^\phl{y \tilde u} D}$$

When $\phl Q$ is of any other shape we can apply the induction hypothesis in a similar (or even simpler) way. \qed

\end{proof}

In the binary case, it is possible to prove a stronger version  of this lemma where $\cut$ is not needed to construct $\phl{R}$, hence its rank is trivially null.

\begin{lemma} 
	\label{lem:cuttwo-adm-simp}
	Let $\phl P\seq\pairQ z{\dual A}, \pairQ xA$ and
	$\phl Q\seq\Gamma, \dbr{\Psi_1}\br[u]{\pairQ w{\dual A}}\dbr{\Psi_2}\pairQ y{B}$.
	Then, there exists $\phl R$ such that
	$\phl R\seq\Gamma, \dbr{\Psi_1}\br[u]{\pairQ z{\dual A}}\dbr{\Psi_2}\pairQ y{B}$.
\end{lemma}

\begin{proof}
	By induction on the structure of $\phl Q$.
	
	Because of the shape of its typing environment, $\phl Q$ cannot be of the shape $\fwd ab$ nor of the shape $\close a$.
	
	When $\phl Q = \send uvST$ and $\phl w\in \mathtt{fn}(\phl S)$, we can take $\phl R = \send uxPT$.
	
	
	When $\phl Q = \send uvST$ and $\phl w\in \mathtt{fn}(\phl T)$, by applying the induction hypothesis to $\phl P$ and $\phl T$ (subprocess of $\phl Q$), we obtain $\phl{R_0}$
	and take $\phl R = \send uvS{R_0}$.
	
	%
	
	When $\phl Q$ is of any other shape we can apply the induction hypothesis in a similar (simpler) way. 
\end{proof}

\subsection{Admissibility of Cut}

\begin{theorem} [Admissibility of Cut]
  Let $\phl P\seq\Gamma_1, \dbr{\Psi_1}\pairQ xA$ and
  $\phl Q\seq\Gamma_2, \dbr{\Psi_2}\pairQ y{A^\perp}$.
  For any $\Gamma$ such that
  $$\big\{\Gamma_1\rhd\dbr{\Psi_1}\pairQ xA,\ \Gamma_2\rhd\dbr{\Psi_2}\pairQ y{\dual
    A}\big\}\longrightarrow^*_{\mathsf{distr}}\longrightarrow^*_{\mathsf{subst}}\Gamma$$
  there exists $\phl R \seq \Gamma$.
\end{theorem}
\begin{proof}
	We proceed by lexicographic induction on the structures of $A$, $\phl P$ and $\phl Q$.
	That is, the induction hypothesis may be applied whenever 
	($i$) the rank of the cut gets smaller,
	or ($ii$) the rank stays the same and the cut is applied to at least one smaller process while the other stays the same.

	\mypar{Axiom}
	
	\[
	\infer[\cut]{
		\phl{\res{xy}(\fwd zx \pp Q)}
		\seq
		\dbr{\Psi_2}\pairQ{z}{\dual a}, \Gamma_2[\phl z/ \phl y]
		}{
		\infer[\tsc{Ax}]{
			\fwd zx\seq\pairQ{z}{\dual a}, \pairQ xa
		}{}
		&
		\phl Q\seq \dbr{\Psi_2}\pairQ y{a^\perp}, \Gamma_2
		}
	\quad\reducesto_\beta\quad
	\phl{Q[z/y]}
	\seq
	\dbr{\Psi_2}\pairQ{z}{\dual a}, \Gamma_2[\phl z/ \phl y]
	\]
	
	\mypar{Units} 
	$$
	\infer[\textsc{Cut}]{
		\phl{\res {xy}(\close x \pp \wait {y}{Q}  )} \seq 
		\Gamma
	}{
		\infer[\one]{
			\close x \seq 
			\set[i]{\br[x]{\Star}\pairQ{u_i}{\cdot}}, \pairQ{x}{\one^\phl{\tilde u}}
		}{}
		&
		\infer[\bot]{
			\wait yQ \seq 
			\dbr{\Psi_2}\pairQ{y}{\bot^\phl{v}}, \Gamma_2,\dbr{\Psi_c}\pairQ{v}{C\ctx{\one^\phl{y\tilde{z}}}}
		}{
			\phl{Q} \seq
			\dbr{\Psi_2}\br[v]{\Star}\pairQ{y}{\cdot}, \Gamma_2,\dbr{\Psi_c}\pairQ{v}{C\ctx{\one^\phl{y\tilde{z}}}}
		}
	}
	$$
	$$
	\quad\reducesto_\beta\quad
	\infer[y \leadsto \tilde u]{
		\phl{Q^\mathsf{y \leadsto \tilde u}}
		\seq
		\set[i]{\dbr{\Psi_{u_i}}\br[v]{\Star}\pairQ{u_i}{\cdot}},
		\Gamma_2^\mathsf{y \leadsto \tilde u},\dbr{\Psi_c}\pairQ{v}{C^\mathsf{y \leadsto \tilde u}\ctx{\one^\phl{\tilde u\tilde{z}}}}
	}{
		\phl{Q} \seq \dbr{\Psi_2}\br[v]{\Star}\pairQ{y}{\cdot}, \Gamma_2, \dbr{\Psi_c}\pairQ{v}{C\ctx{\one^\phl{y\tilde{z}}}}
	}
	$$
	
	This is a special rule: It is not a uniform substitution replacing $\phl y$ by $\phl {\tilde{u}}$ but it reproduces the changes performed by the distribution of $\Psi$ over the $\phl{u_i}$s.
	Because $\dbr\Psi$ is distributed from a terminated endpoint $\phl y$ to another list of endpoints in $\phl{\tilde u}$ that are also terminated, this rule is trivially admissible.

	$$
	\left\{
	\begin{array}{rcl}
		\set[i]{\br[x]{\Star}\pairQ{u_i}{\cdot}} & \rhd & \pairQ{x}{\one^{\tilde u}}\\
		\Gamma_2,\dbr{\Psi_c}\pairQ{v}{C\ctx{\one^\phl{\tilde{z_1}{y}\tilde{z_2}}}} & \rhd & \dbr{\Psi}\pairQ{y}{\bot^v}
	\end{array}
	\right\}
	\xrightarrow{}_{\mathsf{distr}}^*
	\left\{
	\begin{array}{rcl}
		\set[i]{\dbr{\Psi_{u_i}}\br[x]{\Star}\pairQ{u_i}{\cdot}} & \rhd & \pairQ{x}{\one^{\tilde u}}\\
		\Gamma_2^\mathsf{y \leadsto \tilde u},\dbr{\Psi_c}\pairQ{v}{C^\mathsf{y \leadsto \tilde u}\ctx{\one^\phl{\tilde{z_1}{y}\tilde{z_2}}}} & \rhd & \pairQ{y}{\bot^v}
	\end{array}
	\right\}
	$$
	$$
	\xrightarrow{}_{\mathsf{subst}}
	\set[i]{\dbr{\Psi_{u_i}}\br[v]{\Star}\pairQ{u_i}{\cdot}},
	\Gamma_2^\mathsf{y \leadsto \tilde u},\dbr{\Psi_c}\pairQ{v}{C^\mathsf{y \leadsto \tilde u}\ctx{\one^\phl{\tilde{z_1}{\tilde u}\tilde{z_2}}}}
	= \Gamma
	$$

	%

	\mypar{Multiplicatives}
	$$
	\infer[\textsc{Cut}]{
		\phl{\res{xy}(\send xa{P}{Q} \pp \recv yc{R}) }
		\seq
		\Gamma
	}{
		\infer[\tensor]{
			\phl{\send xa{P}{Q}} 
			\seq 
			\Gamma_1,
			\br[x]{\pairQ{d}{\dual A}}\dbr{\Psi_u}\pairQ{u}{C},
			\dbr{\Psi_1}\pairQ x{A \tensor^\phl{u} B}
		}{
			\phl {P}
			\seq
			\pairQ{d}{\dual A},\pairQ aA
			& 
			\phl {Q} 
			\seq
			\Gamma_1, \dbr{\Psi_u}\pairQ{u}{C}, \dbr{\Psi_1}\pairQ xB
		}
		&
		\infer[\parr]{
			\phl{\recv yc{R}} 
			\seq 
			\dbr{\Psi_2}\pairQ y{\dual A \parr^v \dual B}, \Gamma_2, \dbr{\Psi_v}\pairQ{v}{D\ctx{\tensor^\phl{y}}}
		}{
			\phl {R}
			\seq 
			\dbr{\Psi_2}\br[v]{\pairQ c{\dual A}}\pairQ y{\dual B}, \Gamma_2, \dbr{\Psi_v}\pairQ{v}{D\ctx{\tensor^\phl{y}}}
		}
	}
	$$
	
	$$
	\quad\reducesto_\beta\quad
	\infer[\tsc{Cut}]{
		\phl {\res{xy} (Q\pp \res{a\rhd c} (P\pp R))}
		\seq
		\Gamma
	}{
		\phl {Q} 
		\seq
		\Gamma_1, \dbr{\Psi_u}\pairQ{u}{C}, \dbr{\Psi_1}\pairQ xB
		&
		\phl {\res{a\rhd c} (P\pp R)}
		\seq 
		\dbr{\Psi_2}\br[v]{\pairQ{d}{\dual A}}\pairQ y{\dual B}, \Gamma_2, \dbr{\Psi_v}\pairQ{v}{D\ctx{\tensor^\phl{y}}}
	}
	$$
	
	By Lemma~\ref{lem:cuttwo-adm-simp}, we know that $\phl S \seq \dbr{\Psi_2}\br[v]{\pairQ{d}{\dual A}}\pairQ y{\dual B}, \Gamma_2, \dbr{\Psi_v}\pairQ{v}{D\ctx{\tensor^\phl{y}}}$, which gives us the right premiss above.
	Then, the new $\cut$ (on $B$) is admissible by induction hypothesis on the rank as $\mathtt{size}(B) < \mathtt{size}(A \tensor^u B)$.
	
	$$
	\left\{
	\begin{array}{rcl}
		\Gamma_1,
		\br[x]{\pairQ{d}{\dual A}}\dbr{\Psi_u}\pairQ{u}{C} & \rhd & 
		\dbr{\Psi_1}\pairQ x{A \tensor^\phl{u} B}\\
		\Gamma_2, \dbr{\Psi_v}\pairQ{v}{D\ctx{\tensor^\phl{y}}} & \rhd &  \dbr{\Psi_2}\pairQ y{\dual A \parr^\phl{v} \dual B} 
	\end{array}
	\right\}
	%
	%
	\xrightarrow{}_{\mathsf{distr}}^*
	\left\{
	\begin{array}{rcl}
		\Gamma_1'',
		\dbr{\Psi_{2u}}\br[x]{\pairQ{d}{\dual A}}\dbr{\Psi_u}\pairQ{u}{C''} & \rhd & 
		\pairQ x{A \tensor^\phl{u} B}\\
		\Gamma_2'', \dbr{\Psi_{1v}}\dbr{\Psi_v}\pairQ{v}{D''\ctx{\tensor^\phl{y}}} & \rhd &  \pairQ y{\dual A \parr^\phl{v} \dual B} 
	\end{array}
	\right\}
	$$
	
	$$
	\xrightarrow{}_{\mathsf{subst}}
	\left\{
	\begin{array}{rcl}
		\Gamma_1'',
		\dbr{\Psi_{2u}}\br[v]{\pairQ{d}{\dual A}}\dbr{\Psi_u}\pairQ{u}{C''} & \rhd & 
		\pairQ x{B}\\
		\Gamma_2'', \dbr{\Psi_{1v}}\dbr{\Psi_v}\pairQ{v}{D''\ctx{\tensor^\phl{u}}} & \rhd &  \pairQ y{\dual B} 
	\end{array}
	\right\}
	\xrightarrow{}_{\mathsf{subst}}^*
	\Gamma
	$$
	
	On the other hand:
	$$
	\left\{
	\begin{array}{rcl}
		\Gamma_1, \dbr{\Psi_u}\pairQ{u}{C} & \rhd &
		\dbr{\Psi_1}\pairQ x{B}\\
		\Gamma_2, \dbr{\Psi_v}\pairQ{v}{D\ctx{\tensor^\phl{y}}} & \rhd & \dbr{\Psi_2}\br[v]{\pairQ{d}{\dual A}}\pairQ y{\dual B}
	\end{array}
	\right\}
	\xrightarrow{}_{\mathsf{distr}}^*
	\left\{
	\begin{array}{rcl}
		\Gamma_1'',
		\dbr{\Psi_{2u}}\bullet\dbr{\Psi_u}\pairQ{u}{C''} & \rhd & 
		\pairQ x{B}\\
		\Gamma_2'', \dbr{\Psi_{1v}}\bullet\dbr{\Psi_v}\pairQ{v}{D''\ctx{\tensor^\phl{y}}} & \rhd &  \br[v]{\pairQ{d}{\dual A}}\pairQ y{\dual B} 
	\end{array}
	\right\}
	$$
	$$
	\xrightarrow{}_{\mathsf{distr}}
	\left\{
	\begin{array}{rcl}
		\Gamma_1'',
		\dbr{\Psi_{2u}}\br[v]{\pairQ{d}{\dual A}}\dbr{\Psi_u}\pairQ{u}{C''} & \rhd & 
		\pairQ x{B}\\
		\Gamma_2'', \dbr{\Psi_{1v}}\dbr{\Psi_v}\pairQ{v}{D''\ctx{\tensor^\phl{u}}} & \rhd &  \pairQ y{\dual B} 
	\end{array}
	\right\}
	\xrightarrow{}_{\mathsf{subst}}^*
	\Gamma
	$$
	
\mypar{Multiplicatives with gathering}

\[
\infer[\textsc{Cut}]{
	\phl{\res{xy}(\send xa{P}{Q} \pp \recv yc{R}) }
	\seq
	\Gamma
}{
	\infer[\tensor]{
		\phl{\send xa{P}{Q}} 
		\seq 
		\Gamma_1,
		\big\{\br[\phl x]{\Delta_i}\dbr{\Psi_{u_i}}\pairQ{u_i}{C_i}\big\}_i,
		\dbr{\Psi_1}\pairQ x{A \tensor^\phl{\tilde u} B}
	}{
		\phl {P}
		\seq
		\big\{\Delta_i\big\}_i, \pairQ aA
		& 
		\phl {Q} 
		\seq
		\Gamma_1, \big\{\dbr{\Psi_{u_i}}\pairQ{u_i}{C_i}\big\}_i, \dbr{\Psi_1}\pairQ xB
	}
	&
	\infer[\parr]{
		\phl{\recv yc{R}} 
		\seq 
		\dbr{\Psi_2}\pairQ y{\dual A \parr^v \dual B}, \Gamma_2, \dbr{\Psi_v}\pairQ{v}{D\ctx{\tensor^\phl{y\tilde z}}}
	}{
		\phl {R}
		\seq 
		\dbr{\Psi_2}\br[v]{\pairQ c{\dual A}}\pairQ y{\dual B}, \Gamma_2, \dbr{\Psi_v}\pairQ{v}{D\ctx{\tensor^\phl{y\tilde z}}}
	}
}
\]

With $\Gamma$ being defined as one of the possible rewriting:
\[
\left\{
\begin{array}{ll}
	\Gamma_1,
	\big\{\br[\phl x]{\Delta_i}\dbr{\Psi_{u_i}}\pairQ{u_i}{C_i}\big\}_i
	\rhd 
	\dbr{\Psi_1}\pairQ x{A \tensor^\phl{\tilde u} B}
	\\
	\Gamma_2, \dbr{\Psi_v}\pairQ{v}{D\ctx{\tensor^\phl{y\tilde z}}} 
	\rhd 
	\dbr{\Psi_2}\pairQ y{\dual A \parr^\phl{v} \dual B}
\end{array}
\right\}
\longrightarrow^*_{\mathsf{distr}}\longrightarrow^*_{\mathsf{subst}}
\Gamma
\]

We want to reduce the cut as follows:
\[
\infer[\tsc{Cut}]{
	\phl {\res{xy} (Q\pp \res{a\rhd c} (P\pp R))}
	\seq
	\Gamma
}{
	\phl {Q} 
	\seq
	\Gamma_1, \big\{\dbr{\Psi_{u_i}}\pairQ{u_i}{C_i}\big\}_i, \dbr{\Psi_1}\pairQ xB
	&
	\phl S 
	\seq 
	\dbr{\Psi_2}\br[v]{\big\{\Delta_i\big\}_i}\pairQ y{\dual B}, \Gamma_2, \dbr{\Psi_v}\pairQ{v}{D\ctx{\tensor^\phl{y\tilde z}}}
}
\]

By Lemma~\ref{lem:cuttwo-adm}, we know that $\phl S = \phl{\res{a\rhd c}(P\pp R)}$ exists, that 
$$\phl S \seq^\cut \dbr{\Psi_2}\br[v]{\big\{\Delta_i\big\}_i}\pairQ y{\dual B}, \Gamma_2, \dbr{\Psi_v}\pairQ{v}{D\ctx{\tensor^\phl{y\tilde z}}}$$
and that its rank is equal to the size of $A$. 
By induction hypothesis any cut in the derivation of $\phl S$ is admissible as $\mathtt{rank}(\phl S) = \mathtt{size}(A) < \mathtt{size}(A \tensor^\phl{\tilde u} B)$.
That means in fact that 
$$\phl S \seq \dbr{\Psi_2}\br[v]{\big\{\Delta_i\big\}_i}\pairQ y{\dual B}, \Gamma_2, \dbr{\Psi_v}\pairQ{v}{D\ctx{\tensor^\phl{y\tilde z}}}$$ 
without the need for cut, which gives us the right premiss.

Then, we need to check that we also get $\Gamma$ as conclusion, namely that we get 
\[
\left\{
\begin{array}{ll}
	\Gamma_1, \big\{\dbr{\Psi_{u_i}}\pairQ{u_i}{C_i}\big\}_i
	& \rhd 
	\dbr{\Psi_1}\pairQ xB
	\\
	\Gamma_2, \dbr{\Psi_v}\pairQ{v}{D\ctx{\tensor^\phl{y\tilde z}}}
	& \rhd 
	\dbr{\Psi_2}\br[v]{\big\{\Delta_i\big\}_i}\pairQ y{\dual B}
\end{array}
\right\}
\longrightarrow^*_{\mathsf{distr}}\longrightarrow^*_{\mathsf{subst}}\Gamma
\]

Finally, the reduced $\cut$ (on $B$) is of course admissible by induction hypothesis on the rank as $\mathtt{size}(B) < \mathtt{size}(A \tensor^\phl{\tilde u} B)$.

\mypar{Additives}
$$
\infer[\tsc{Cut}]{
	\phl{\res{xy}(\Case{x}{P}{Q} \pp \inl yR )}
	\seq \Gamma
}{
	\infer[\with]{
		\Case{x}{P}{Q}
		\seq \Gamma_1, \big\{\dbr{\Psi_{u_i}}\pairQ{u_i}{C_i\ctx{\oplus^\phl{x}}}\big\}_i,
		\dbr{\Psi_1}\pairQ x{A \with^\phl{\tilde u} B}
	}{
		\phl {P}\seq \ldots,
		\dbr{\Psi_1}\br[\tilde u]{\Left}\pairQ xA
		&
		\phl {Q}\seq \ldots,
		\dbr{\Psi_1}\br[\tilde u]{\Right}\pairQ xB
	}
	&
	\infer[\oplus_l]{
		\inl yR \seq
		\dbr{\Psi_2}\pairQ y{\dual A \oplus^\phl{v} \dual B},
		\Gamma_2, \br[y]{\Left}\dbr{\Psi_v} \phl v: \thl D
	}{
		\phl R \seq
		\dbr{\Psi_2}\pairQ y{\dual A},
		\Gamma_2, \dbr{\Psi_v} \phl v: \thl D
	}
}
$$

\[
\left\{
\begin{array}{lcl}
	\Gamma_1, \big\{\dbr{\Psi_{u_i}}\pairQ{u_i}{C_i\ctx{\oplus^\phl{x}}}\big\}_i
	&\rhd &
	\dbr{\Psi_1}\pairQ x{A \with^\phl{\tilde u} B}
	\\
	\Gamma_2, \br[y]{\Left}\dbr{\Psi_v} \phl v: \thl D
	&\rhd &
	\dbr{\Psi_2}\pairQ y{\dual A \oplus^\phl{v} \dual B}
\end{array}
\right\}
\longrightarrow^*_{\mathsf{distr}}
\left\{
\begin{array}{lcl}
	\Gamma'_1, \big\{\dbr{\Psi_{2u_i}}\dbr{\Psi_{u_i}}\pairQ{u_i}{C'_i\ctx{\oplus^\phl{x}}}\big\}_i
	&\rhd &
	\pairQ x{A \with^\phl{\tilde u} B}
	\\
	\Gamma_2', \dbr{\Psi_{1 v}}\br[y]{\Left}\dbr{\Psi_v} \phl v: \thl D'
	&\rhd &
	\pairQ y{\dual A \oplus^\phl{v} \dual B}
\end{array}
\right\}
\]
\[
\longrightarrow_{\mathsf{subst}}
\left\{
\begin{array}{lcl}
	\Gamma'_1, \big\{\dbr{\Psi_{2u_i}}\dbr{\Psi_{u_i}}\pairQ{u_i}{C'_i\ctx{\oplus^\phl{v}}}\big\}_i
	&\rhd &
	\pairQ x{B}
	\\
	\Gamma_2', \dbr{\Psi_{1 v}}\br[\tilde u]{\Left}\dbr{\Psi_v} \phl v: \thl D'
	&\rhd &
	\pairQ y{\dual B}
\end{array}
\right\}
\longrightarrow^*_{\mathsf{subst}}\Gamma
\]

$$
\infer[\tsc{Cut}]{
	\phl{\res{xy}(P \pp R )}
	\seq \Gamma
}{
	\phl {P}\seq \Gamma_1, \big\{\dbr{\Psi_{u_i}}\pairQ{u_i}{C_i\ctx{\oplus^\phl{x}}}\big\}_i,
	\dbr{\Psi_1}\br[\tilde u]{\Left}\pairQ xA
	&
	\phl R \seq
	\dbr{\Psi_2}\pairQ y{\dual A},
	\Gamma_2, \dbr{\Psi_v} \phl v: \thl D
}
$$

\[
\left\{
\begin{array}{lcl}
	\Gamma_1, \big\{\dbr{\Psi_{u_i}}\pairQ{u_i}{C_i\ctx{\oplus^\phl{x}}}\big\}_i
	&\rhd &
	\dbr{\Psi_1}\br[\tilde u]{\Left}\pairQ xA
	\\
	\Gamma_2, \dbr{\Psi_v} \phl v: \thl D
	&\rhd &
	\dbr{\Psi_2}\pairQ y{\dual A}
\end{array}
\right\}
\longrightarrow^*_{\mathsf{distr}}
\left\{
\begin{array}{lcl}
	\Gamma'_1, \big\{\dbr{\Psi_{2u_i}}\bullet\dbr{\Psi_{u_i}}\pairQ{u_i}{C'_i\ctx{\oplus^\phl{x}}}\big\}_i
	&\rhd &
	\br[\tilde u]{\Left}\pairQ xA
	\\
	\Gamma_2', \dbr{\Psi_{1 v}}\bullet\dbr{\Psi_v} \phl v: \thl D'
	&\rhd &
	\pairQ y{\dual A}
\end{array}
\right\}
\]
\[
\longrightarrow_{\mathsf{distr}}
\left\{
\begin{array}{lcl}
	\Gamma'_1, \big\{\dbr{\Psi_{2u_i}}\dbr{\Psi_{u_i}}\pairQ{u_i}{C'_i\ctx{\oplus^\phl{v}}}\big\}_i
	&\rhd &
	\pairQ x{B}
	\\
	\Gamma_2', \dbr{\Psi_{1 v}}\br[\tilde u]{\Left}\dbr{\Psi_v} \phl v: \thl D'
	&\rhd &
	\pairQ y{\dual B}
\end{array}
\right\}
\longrightarrow^*_{\mathsf{subst}}\Gamma
\]

\mypar{Exponentials}

$$
\infer[\cut]{
	\phl{\res{xy}(\srv xaP \pp \client ybQ )}
	\seq \Gamma 
}{
	\infer[!]{
		\srv xaP \seq 
		\set[i]{\pairQ{u_i}{\query^\phl{x} B_i}}[\phl x/\phl a], \pairQ x{\bang^\phl{\tilde u} A} 
	}{
		\phl{P} \seq \set[i]{\pairQ{u_i}{\query^\phl{a} B_i}}, \br[\tilde u]{\Query}\pairQ aA
	}
	&
	\infer[?]{
		\client ybQ \seq 
		\set{\Gamma_2, \br[y]{\Query}\dbr{\Psi_z}\pairQ zC}[\phl y/\phl b], 
		\dbr{\Psi_2}\pairQ y{\query^\phl{z} \dual A}
	}{
		\phl{Q}\seq \Gamma_2, \dbr{\Psi_z}\pairQ zC, \dbr{\Psi_2}\pairQ b{\dual A}
	}
}
$$

\[
\left\{
\begin{array}{lcl}
	\set[i]{\pairQ{u_i}{\query^\phl{x} B_i}}[\phl x/\phl a]
	&\rhd &
	\pairQ x{\bang^\phl{\tilde u} A} 
	\\
	\set{\Gamma_2, \br[y]{\Query}\dbr{\Psi_z}\pairQ zC}[\phl y/\phl b]
	&\rhd &
	\dbr{\Psi_2}\pairQ y{\query^\phl{z} \dual A}
\end{array}
\right\}
\longrightarrow^*_{\mathsf{distr}}
\left\{
\begin{array}{lcl}
	\set[i]{\dbr{\Psi_{2u_i}}\pairQ{u_i}{\query^\phl{x} B'_i}}[\phl x/\phl a]
	&\rhd &
	\pairQ x{\bang^\phl{\tilde u} A} 
	\\
	\set{\Gamma'_2, \br[y]{\Query}\dbr{\Psi_z}\pairQ z{C'}}[\phl y/\phl b]
	&\rhd &
	\pairQ y{\query^\phl{z} \dual A}
\end{array}
\right\}
\]

\[
\longrightarrow_{\mathsf{subst}}
\left\{
\begin{array}{lcl}
	\set[i]{\dbr{\Psi_{2u_i}}\pairQ{u_i}{\query^\phl{z} B'_i}}[\phl x/\phl a]
	&\rhd &
	\pairQ x{A} 
	\\
	\set{\Gamma'_2, \br[\tilde u]{\Query}\dbr{\Psi_z}\pairQ z{C'}}[\phl y/\phl b]
	&\rhd &
	\pairQ y{\dual A}
\end{array}
\right\}
\longrightarrow^*_{\mathsf{subst}}
\Gamma
\]

$$
\quad\reducesto_\beta\quad
\infer[\tsc{Cut}]{
	\seq \Gamma
}{
	\phl{P} \seq 
	\set[i]{\pairQ{u_i}{\query^\phl{a} B_i}}, \br[\tilde u]{\Query}\pairQ aA
	&
	\phl{Q}\seq 
	\Gamma_2, \dbr{\Psi_z}\pairQ zC, \dbr{\Psi_2}\pairQ b{\dual A}
}
$$	
	
\[
\left\{
\begin{array}{lcl}
	\set[i]{\pairQ{u_i}{\query^\phl{a} B_i}}
	&\rhd &
	\br[\tilde u]{\Query}\pairQ aA
	\\
		\Gamma_2, \dbr{\Psi_z}\pairQ zC
	&\rhd &
	\dbr{\Psi_2}\pairQ b{\dual A}
\end{array}
\right\}
\longrightarrow^*_{\mathsf{distr}}
\left\{
\begin{array}{lcl}
	\set[i]{\dbr{\Psi_{2u_i}}\pairQ{u_i}{\query^\phl{a} B'_i}}
	&\rhd &
	\br[\tilde u]{\Query}\pairQ aA
	\\
	\Gamma'_2, \dbr{\Psi_z}\pairQ z{C'}
	&\rhd &
	\pairQ b{\dual A}
\end{array}
\right\}
\]
\[
\longrightarrow_{\mathsf{distr}}
\left\{
\begin{array}{lcl}
	\set[i]{\dbr{\Psi_{2u_i}}\pairQ{u_i}{\query^\phl{z} B'_i}}
	&\rhd &
	\pairQ aA
	\\
	\Gamma'_2, \br[\tilde u]{\Query}\dbr{\Psi_z}\pairQ z{C'}
	&\rhd &
	\pairQ b{\dual A}
\end{array}
\right\}
\longrightarrow^*_{\mathsf{subst}}
\Gamma
\]

\mypar{Commutative $\tensor$ case}

\[
\infer[\cut]{
	\phl{\res {xy}(P\pp (\send z{v}{Q}{R}))}\seq 
	\Gamma 
}{
	\phl P\seq\Gamma_1, \set[j]{\dbr{\Psi_{w_j}}\pairQ{w_j}{E_j}}, \dbr{\Psi_1}\pairQ x{A}
	&
	\infer[\tensor]{
		\send z{v}{Q}{R}\seq 
		\Gamma_2,
		\{\br[z]{\Delta_i}\dbr{\Psi_{u_i}}\pairQ{u_i}{C_i}\}_i,
		\br[z]{\Sigma}\dbr{\Psi_2}\pairQ y{\dual A},
		\dbr{\Psi_z}\pairQ z{B\tensor^{\phl{\tilde u y}} D}
	}{
		\phl {Q}\seq \{\Delta_i\}_i, \Sigma, \pairQ {v} {B}
		&
		\phl {R}\seq 
		\Gamma_2,
		\{\dbr{\Psi_{u_i}}\pairQ{u_i}{C_i}\}_i,
		\dbr{\Psi_2}\pairQ y{\dual A}, \dbr{\Psi_z}\pairQ z{D}
	}
}
\]

\[
\left\{
\begin{array}{lcl}
	\Gamma_1, \set[j]{\dbr{\Psi_{w_j}}\pairQ{w_j}{E_j}}
	&\rhd &
	\dbr{\Psi_1}\pairQ x{A}
	\\
	\Gamma_2,
	\{\br[z]{\Delta_i}\dbr{\Psi_{u_i}}\pairQ{u_i}{C_i}\}_i, \dbr{\Psi_z}\pairQ z{B\tensor^{\phl{\tilde u y}} D}
	&\rhd& 
	\br[z]{\Sigma}\dbr{\Psi_2}\pairQ y{\dual A}
\end{array}
\right\}
\]
\[
\longrightarrow_{\mathsf{distr}}
\left\{
\begin{array}{lcl}
	\Gamma_1, \set[j]{\br[z]{\Sigma_j}\bullet\dbr{\Psi_{w_j}}\pairQ{w_j}{E_j}}
	&\rhd &
	\dbr{\Psi_1}\pairQ x{A}
	\\
	\Gamma_2,
	\{\bullet\br[z]{\Delta_i}\dbr{\Psi_{u_i}}\pairQ{u_i}{C_i}\}_i, \bullet\dbr{\Psi_z}\pairQ z{B\tensor^{\phl{\tilde u \tilde w}} D}
	&\rhd& 
	\dbr{\Psi_2}\pairQ y{\dual A}
\end{array}
\right\}
\quad\text{with}\quad\Sigma = \set[j]{\Sigma_j}
\]
\[
\longrightarrow^*_{\mathsf{distr}}
\left\{
\begin{array}{lcl}
	\Gamma'_1, \set[j]{\br[z]{\Sigma_j}\dbr{\Psi_{2w_j}}\dbr{\Psi_{w_j}}\pairQ{w_j}{E'_j}}
	&\rhd &
	\pairQ x{A}
	\\
	\Gamma'_2,
	\{\dbr{\Psi_{1u_i}}\br[z]{\Delta_i}\dbr{\Psi_{u_i}}\pairQ{u_i}{C'_i}\}_i, \dbr{\Psi_{1z}}\dbr{\Psi_z}\pairQ z{B\tensor^{\phl{\tilde u \tilde w}} D'}
	&\rhd& 
	\pairQ y{\dual A}
\end{array}
\right\}
\longrightarrow^*_{\mathsf{subst}}\Gamma
\]

We want to cut $\phl P$ and $\phl R$ together, as we can ultimately apply the induction hypothesis to this $\cut$ given that $\phl R$ is a subprocess of $\send z{v}{Q}{R}$:
\[
\infer[\cut]{
	\phl{\res{xy}(P \pp R)}
	\seq \Gamma_0
}{
	\phl P\seq\Gamma_1, \set[j]{\dbr{\Psi_{w_j}}\pairQ{w_j}{E_j}}, \dbr{\Psi_1}\pairQ x{A}
	&
	\phl {R}\seq 
	\Gamma_2,
	\{\dbr{\Psi_{u_i}}\pairQ{u_i}{C_i}\}_i,
	\dbr{\Psi_2}\pairQ y{\dual A}, \dbr{\Psi_z}\pairQ z{D}
}\]

This gives us conclusion $\Gamma_0$ obtained by reproducing the $\longrightarrow^*_{\mathsf{distr}}$ rewriting sequence on $\dbr{\Psi_1}$ and $\dbr{\Psi_2}$ that we had above, followed by the deterministic $\longrightarrow^*_{\mathsf{subst}}$ sequence on $\pairQ{x}{A}$ and $\pairQ{y}{\dual A}$.

\[
\left\{
\begin{array}{lcl}
	\Gamma_1, \set[j]{\dbr{\Psi_{w_j}}\pairQ{w_j}{E_j}}
	&\rhd& 
	\dbr{\Psi_1}\pairQ x{A}
	\\
	\Gamma_2,\{\dbr{\Psi_{u_i}}\pairQ{u_i}{C_i}\}_i, \dbr{\Psi_z}\pairQ z{D}
	&\rhd& 
	\dbr{\Psi_2}\pairQ y{\dual A}
\end{array}
\right\}
\]
\[
\longrightarrow^*_{\mathsf{distr}}
\left\{
\begin{array}{lcl}
	\Gamma'_1, \set[j]{\dbr{\Psi_{2w_j}}\dbr{\Psi_{w_j}}\pairQ{w_j}{E'_j}}
	&\rhd &
	\pairQ x{A}
	\\
	\Gamma'_2,
	\{\dbr{\Psi_{1u_i}}\dbr{\Psi_{u_i}}\pairQ{u_i}{C'_i}\}_i, \dbr{\Psi_{1z}}\dbr{\Psi_z}\pairQ z{D'}
	&\rhd& 
	\pairQ y{\dual A}
\end{array}
\right\}
\longrightarrow^*_{\mathsf{subst}}\Gamma_0
\]

Finally we can reconstruct the original conclusion $\Gamma$ by applying the $\tensor$ rule.

\[
\infer[\tensor]{
	\send zvQ{\res{xy}(P \pp R)}
	\seq \Gamma
}{
	\phl {Q}\seq \{\Delta_i\}_i, \Sigma, \pairQ {v} {B}
	&
	\phl{\res{xy}(P \pp R)}
	\seq \Gamma_0
}
\]

\mypar{Commutative $\parr$ case}

\[
\infer[\cut]{
	\phl {\res{xy} (P\pp \recv uvQ)}\seq \Gamma
}{
	\phl P\seq
	\Gamma_1, \dbr{\Psi_1}\pairQ x{A}
	&
	\infer[\parr]{
		\phl {\recv uvQ}\seq 
		\Gamma_2, \dbr{\Psi_u}\pairQ u{B\parr^{\phl z}C}, \dbr{\Psi_2}\pairQ{y} {\dual A}
	}{
		\phl {Q}\seq \Gamma_2, \dbr{\Psi_u}\br[z]{\pairQ v{B}}\pairQ u{C}, \dbr{\Psi_2}\pairQ{y} {\dual A}
	}
}
\]

\[
\left\{
\begin{array}{lcl}
	\Gamma_1
	&\rhd& 
	\dbr{\Psi_1}\pairQ x{A}
	\\
	\Gamma_2, \dbr{\Psi_u}\pairQ u{B\parr^{\phl z}C}  
	&\rhd& 
	\dbr{\Psi_2}\pairQ{y} {\dual A}
\end{array}
\right\}
\longrightarrow^*_{\mathsf{distr}}\longrightarrow^*_{\mathsf{subst}}\Gamma
\]

\[
\infer[\parr]{
	\recv uv{\res {xy}(P\pp Q)} \seq\Gamma
}{
	\infer[\cut]{
		\phl{\res {xy}(P\pp Q)} \seq \Gamma_0
	}{
		\phl P\seq
		\Gamma_1, \dbr{\Psi_1}\pairQ x{A}
		&
		\phl {Q}\seq \Gamma_2, \dbr{\Psi_u}\br[z]{\pairQ v{B}}\pairQ u{C}, \dbr{\Psi_2}\pairQ{y} {\dual A}
	}
}
\]

\[
\left\{
\begin{array}{lcl}
	\Gamma_1
	&\rhd& 
	\dbr{\Psi_1}\pairQ x{A}
	\\
	\Gamma_2, \dbr{\Psi_u}\br[z]{\pairQ v{B}}\pairQ u{C}
	&\rhd& 
	\dbr{\Psi_2}\pairQ{y} {\dual A}
\end{array}
\right\}
\longrightarrow^*_{\mathsf{distr}}\longrightarrow^*_{\mathsf{subst}}\Gamma_0
\]

\mypar{Commutative $\bot$ case}

\[
\infer[\cut]{
	\phl{\res{xy}(P \pp \wait zQ)}
	\seq
	\Gamma
}{
	\phl{P} \seq \Gamma_1, \dbr{\Psi_1}\pairQ{x}{A}
	&
	\infer[\bot]{
		\wait zQ \seq 
		\dbr{\Psi_2}\pairQ{y}{\dual A}, \Gamma_2,
		\dbr{\Psi_z}\pairQ{z}{\bot^\phl{v}},
		\dbr{\Psi_v}\pairQ{v}{C\ctx{\one^\phl{z\tilde{u}}}}
	}{
		\phl{Q} \seq
		\dbr{\Psi_2}\pairQ{y}{\dual A}, \Gamma_2,
		\dbr{\Psi_z}\br[v]{\Star}\pairQ{z}{\cdot}, \dbr{\Psi_v}\pairQ{v}{C\ctx{\one^\phl{z\tilde{u}}}}
	}
}
\]

\[
\left\{
\begin{array}{l@{\quad\rhd\quad}l}
	\Gamma_1
	&
	\dbr{\Psi_1}\pairQ x{A}
	\\
	\Gamma_2,\dbr{\Psi_z}\pairQ{z}{\bot^\phl{v}},
	\dbr{\Psi_v}\pairQ{v}{C\ctx{\one^\phl{z\tilde{u}}}}
	&
	\dbr{\Psi_2}\pairQ{y}{\dual A}
\end{array}
\right\}
\longrightarrow^*_{\mathsf{distr}}\longrightarrow^*_{\mathsf{subst}}
\Gamma
\]

\[
\infer[\cut]{
	\phl{\res{xy}(P \pp Q)}
	\seq
	\Gamma_0
}{
	\phl{P} \seq \Gamma_1, \dbr{\Psi_1}\pairQ{x}{A}
	&
	\phl{Q} \seq
	\dbr{\Psi_2}\pairQ{y}{\dual A}, \Gamma_2,
	\dbr{\Psi_z}\br[v]{\Star}\pairQ{z}{\cdot}, \dbr{\Psi_v}\pairQ{v}{C\ctx{\one^\phl{z\tilde{u}}}}
}
\]

\[
\left\{
\begin{array}{l@{\quad\rhd\quad}l}
	\Gamma_1
	&
	\dbr{\Psi_1}\pairQ x{A}
	\\
	\Gamma_2,
	\dbr{\Psi_z}\br[v]{\Star}\pairQ{z}{\cdot}, \dbr{\Psi_v}\pairQ{v}{C\ctx{\one^\phl{z\tilde{u}}}}
	&
	\dbr{\Psi_2}\pairQ{y}{\dual A}
\end{array}
\right\}
\longrightarrow^*_{\mathsf{distr}}\longrightarrow^*_{\mathsf{subst}}
\Gamma
\]

\[
\infer[\bot]{
	\wait{z}{ \phl{\res{xy}(P \pp Q)}}
	\seq \Gamma
}{
	\phl{\res{xy}(P \pp Q)}
	\seq
	\Gamma_0
}
\]
	
\end{proof}

\section{MCutF admissibility}\label{app:generalise}
Here are all cases for \tsc{MCutF} elimination.

\begin{displaymath}\small
  \infer[\tsc{MCutF}]{
  	\phl{\res{\tilde x: \res{\tilde y}Q \br{\tilde P}} \tilde R} \cll \tilde\Delta, \tilde\Sigma
  }{
  	\set[j\le m]{\phl {P_j} \cll \thl\Delta_i, \pairQ{y_j}{A_j}}
    &
    \set[i\le n]{\phl {R_i} \cll \thl\Sigma_i, \pairQ{x_i}{B_i}}
    &
    \phl Q \seq \set[i\le n]{\dbr{\Psi_i}\pairQ{x_i}\dual{B_i}}, \set[n < i \le p]{ \dbr{\Psi_i}\pairQ{x_i}{\cdot}}
  }
\end{displaymath}
The rule has a developed side condition: $\bigcup_{i\le p}{\Psi_i} \setminus
\set{\Left, \Right, \Query, \Star} = \set[j\le
m]{\pairQ{y_j}{\dual{A_j}}}$.

\mypar {Send Message ($\tensor$).}
\begin{displaymath}\small
  \infer[\tsc{MCutF}]{
  	\phl{\res{x\tilde x:\res{\tilde y}\recv xyQ\br{\tilde P}}(\send{x}{y}{P}{R} \pp \tilde R)} \cll \Delta, \Sigma, \tilde\Delta, \tilde\Sigma
  }
  {
    \infer[\tensor]
    {
      \phl{\send{x}{y}{P}{R}}\cll \Delta,\Sigma, \pairQ x{A\tensor B}
    }
    {
      \phl P\cll \Delta,\pairQ yA
      &
      \phl R\cll \Sigma,\pairQ xB
    }
    & \quad\ldots\quad
    &
    \infer[\parr]
    {
      \phl{\recv xyQ} \seq 
      \dbr{\Psi}\pairQ x{\dual A\parr^{\phl{x_k}}\dual B}, 
      \Gamma
    }
    {
      \phl Q\seq \dbr{\Psi}\br[x_k]{\pairQ y{\dual A}}\pairQ x{\dual B}, 
      \Gamma
    }
  }
\end{displaymath}
\qquad\qquad$\Longrightarrow$ 
\begin{displaymath}\small
  \infer[\tsc{MCutF}]{
  	\phl{\res{x\tilde x:\res{y\tilde y}Q\br{P, \tilde P}}(R \pp \tilde R)} \cll \Delta, \Sigma, \tilde\Delta, \tilde\Sigma
  }
  {
    	\phl P\cll\Delta,\pairQ yA
    	&
    	\phl R\cll\Sigma,\pairQ xB
    & \quad\ldots\quad
    &
    \phl Q\seq 
    \dbr{\Psi}\br[x_k]{\pairQ y\dual{A}}\pairQ x\dual{B}, 
    \Gamma
  }
\end{displaymath}

\mypar{Receive Message ($\parr$).} 
\begin{displaymath}\small
  \infer[\tsc{MCutF}]
  {\phl{\res{x\tilde x:\res{z\tilde y}\send xy{S}Q\br{P,\tilde P}}(\recv{x}{y}R \pp \tilde R)}
  	\cll 
  	\Delta, \Sigma, \tilde\Delta, \tilde\Sigma
  }
  {
    \begin{array}{ll}
      \phl{P} \cll \Delta, \pairQ{z}\dual{A}
      \\[2mm]
      \infer[\parr]
      {
      \phl{\recv{x}{y}{R}}\cll \Sigma, \pairQ x{A\parr B}
      }
      {
      \phl{R}\cll \Sigma, \pairQ yA, \pairQ x{B}
      }
    \end{array}
    &
    \ldots
    &
    \infer[\tensor]{
      \phl{\send xy{S}Q} 
      \seq 
      \dbr{\Psi_x}\pairQ x{\dual A \tensor^{\phl{x_k}} \dual B}, \br[x]{\pairQ{z}{A}}\dbr{\Psi_k}\pairQ{x_k}\dual{B_k},
      \Gamma-k
    }{
    	\phl S
    	\seq
    	\pairQ{z}{A}, \pairQ y{\dual A}
    	&
      \phl Q
      \seq
	\dbr{\Psi_x}\pairQ x{\dual B}, 
	\Gamma
    }
  }
\end{displaymath}
\qquad\qquad$\Longrightarrow$ 
\begin{displaymath}\small
  \infer[\tsc{MCutF}]
  {\phl{\res{x\tilde x:\res{\tilde y}Q\br{\tilde P}}(\res{yz:S}(R\pp P) \pp \tilde R)} \cll 
  \Delta, \Sigma, \tilde\Delta, \tilde\Sigma
	}
  {
    \infer[\tsc{MCutF}]
    {
      \phl{\res{yz:S}(R\pp P)}\seq\Sigma, \Delta, \pairQ xB
    }
    {
      \phl R\cll \Sigma, \pairQ yA, \pairQ x{B}
      &
      \phl P\cll\Delta,\pairQ z{\dual A}
      &
      \phl S\seq\pairQ zA,\pairQ y{\dual A}
    }
    &\quad\ldots\quad
    &
    \phl Q
    \seq
    \dbr{\Psi_x}\pairQ x{\dual B}, 
    \Gamma
  }
\end{displaymath}
Note that now the message (namely process $\phl P$) has finally been
delivered and it can be directly linked with a new \tsc{MCutF}.

\mypar{Internal choice ($\oplus$).} 
\begin{displaymath}\small
	\infer[\tsc{MCutF}]
	{\phl{\res{x\tilde x:\res{\tilde y}\Case x{Q}S\br{\tilde P}}(\inl{x}{R} \pp \tilde R)}
		\cll 
		\Sigma, \tilde\Delta, \tilde\Sigma
	}
	{
		\infer[\oplus_l]
		{
			\phl{\inl{x}{R}}\cll \Sigma, \pairQ x{A\oplus B}
		}
		{
			\phl{R}\cll \Sigma, \pairQ x A
		}
		&\quad\ldots\quad
		&
		\infer[\with]{
			\phl{\Case x{Q}S} 
			\seq
			\dbr{\Psi_x}\pairQ x{\dual A \with^{\phl{x_k}} \dual B}, 
			\Gamma
		}{
			\phl Q
			\seq
			\dbr{\Psi_x}\br[x_k]{\Left}\pairQ x{\dual A}, 
			\Gamma
			&
			\phl S
			\seq
			\dbr{\Psi_x}\br[x_k]{\Right}\pairQ x{\dual B}, 
			\Gamma
		}
	}
\end{displaymath}
\qquad\qquad$\Longrightarrow$ 
\begin{displaymath}\small
	\infer[\tsc{MCutF}]
	{\phl{\res{x\tilde x:\res{\tilde y}Q\br{\tilde P}}(R \pp \tilde R)}
		\cll 
		\Sigma, \tilde\Delta, \tilde\Sigma
	}
	{
		\phl{R}\cll \Sigma, \pairQ x A
		&\quad\ldots\quad
		&
			\phl Q
			\seq
			\dbr{\Psi_x}\br[x_k]{\Left}\pairQ x{\dual A}, 
			\Gamma
	}
\end{displaymath}

\mypar{External choice ($\with$).} 
\begin{displaymath}\small
	\infer[\tsc{MCutF}]{
		\phl{\res{x\tilde x:\res{\tilde y}\inl{x}{Q}\br{\tilde P}}(\Case{x}{R}{S} \pp \tilde R)} \cll \Sigma, \tilde\Delta, \tilde\Sigma
	}
	{
		\infer[\with]
		{
			\phl{\Case{x}{R}{S}}\cll \Sigma,\pairQ x{A\with B}
		}
		{
			\phl R\cll \Sigma,\pairQ yA
			&
			\phl S\cll \Sigma,\pairQ xB
		}
		& \quad\ldots\quad
		&
		\infer[\oplus_l]{
			\inl xQ\seq 
			\dbr{\Psi_x}\pairQ x{\dual A \oplus^{x_k} \dual B}, \br[x]{\Left}\dbr{\Psi_k} \phl {x_k}: \thl \dual{B_k},
			\Gamma-k
		}{
		\phl Q\seq 
			\dbr{\Psi_x}\pairQ x{\dual A}, 
			\Gamma
		}
	}
\end{displaymath}
\qquad\qquad$\Longrightarrow$
\begin{displaymath}\small
	\infer[\tsc{MCutF}]{
		\phl{\res{x\tilde x:\res{\tilde y}{Q}\br{\tilde P}}({R} \pp \tilde R)} \cll \Sigma, \tilde\Delta, \tilde\Sigma
	}
	{
			\phl R\cll \Sigma,\pairQ yA
		&\quad\ldots\quad
		&
			\phl Q\seq 
			\dbr{\Psi_x}\pairQ x{\dual A}, 
			\Gamma
	}
\end{displaymath}

\mypar{Axiom.} 
%
\begin{displaymath}\small
	\infer[\tsc{MCutF}]{
		\phl{\res{x_1x_2:\fwd {x_1}{x_2}}(\fwd {x_1}{z_1} \pp \fwd {x_2}{z_2})} \cll \pairQ {z_1}{\dual a}, \pairQ {z_2}{a}
	}
	{
		\infer[\tsc{Ax}]{
			\fwd {x_1}{z_1}\cll\pairQ{x_1}{a}, \pairQ {z_1}{\dual a}
		}{}
		&
		\infer[\tsc{Ax}]{
			\fwd {x_2}{z_2} \cll\pairQ{x_2}{\dual a}, \pairQ {z_2}{a}
		}{}
		&
		\infer[\tsc{Ax}]{
			\fwd {x_1}{x_2} \seq\pairQ{x_1}{\dual a}, \pairQ {x_2}{a}
		}{}
	}
\end{displaymath}
\qquad\qquad$\Longrightarrow$
\begin{displaymath}\small
	\infer[\tsc{Ax}]{
		\fwd {z_1}{z_2}\cll\pairQ {z_1}{\dual a}, \pairQ {z_2}{a}
	}{}
\end{displaymath}

\mypar{Close ($\one$).}
\begin{displaymath}\small
	\infer[\tsc{MCutF}]{
		\phl{\res{x\tilde x:\res{\tilde y}\wait xQ\br{\tilde P}}(\close x \pp \tilde R)} \cll \tilde\Delta, \tilde\Sigma
	}
	{
		\infer[\one]
		{\phl{\close x} \cll \phl x: \thl\one}
		{ }
		&\quad\ldots\quad
		&
		\infer[\bot]{
			\wait xQ \seq 
			\dbr{\Psi_x}\pairQ{x}{\bot^{x_k}}, 
			\Gamma
		}{
			\phl{Q} \seq 
			\dbr{\Psi_x}\br[x_k]{\Star}\pairQ{x}{\cdot}, 
			\Gamma
		}
	}
\end{displaymath}
\qquad\qquad$\Longrightarrow$
\begin{displaymath}\small
	\infer[\tsc{MCutF}]{
		\phl{\res{x\tilde x:\res{\tilde y}Q\br{\tilde P}}(\tilde R)} \cll \tilde\Delta, \tilde\Sigma
	}
	{
		\quad\ldots\quad
		&
			\phl{Q} \seq 
			\dbr{\Psi_x}\br[x_k]{\Star}\pairQ{x}{\cdot}, 
			\Gamma
	}
\end{displaymath}

\mypar{Wait ($\bot$).}
\begin{displaymath}\small
	\infer[\tsc{MCutF}]{
		\phl{\res{x:\close x }(\wait xP)} \cll \Delta
	}
	{
		\infer[\bot]
		{\phl {\wait xP} \cll \thl \Delta, \phl x:\thl\bot}
		{\phl P \cll \thl\Delta}
		&
		\infer[\one\ (\vec u \not= \emptyset)]{
			\close x \seq \pairQ{x}{\one^{\tilde x}}, \set[i]{\br[x]{\Star}\pairQ{x_i}{\cdot}}
		}{ }
	}
\end{displaymath}
\qquad\qquad$\Longrightarrow$\quad
${\phl P \cll \thl\Delta}$

\begin{theorem}[Admissibility of multi-cut]
  If
  $\set[j\le m]{\phl {P_j} \cll \thl\Delta_i, \pairQ{y_j}{A_j}}$
  and
  $\set[i\le n]{\phl {R_i} \cll \thl\Sigma_i, \pairQ{x_i}{B_i}}$
  and 
  $\phl Q \seq \set[i\le n]{\dbr{\Psi_i}\pairQ{x_i}\dual{B_i}}, \set[n < i \le p]{ \dbr{\Psi_i}\pairQ{x_i}{\cdot}}$
  then  
  $\phl{\res{\tilde x: \res{\tilde y}Q \br{\tilde P}} \tilde R} \cll \tilde\Delta, \tilde\Sigma$
\end{theorem}


\end{document}